\documentclass[%
 reprint,
 amsmath,amssymb,
 aps,
 prx,
]{revtex4-2}

\usepackage{graphicx}
\usepackage{dcolumn}
\usepackage{bm}
\usepackage{tikz-feynman}
\usepackage{makecell}

\newcommand{\llrrangle}[1]{
\langle\mkern-4mu \langle #1\rangle \mkern-4mu \rangle}

\begin{document}


\title{Republished: Dynamics of stochastic integrate-and-fire networks}

\author{Gabriel Koch Ocker}%
 \email{gkocker@bu.edu}
\affiliation{%
Department of Mathematics and Statistics\\
Boston University\\ 
Boston, MA 02215
}%

%
%

\date{\today}

\begin{abstract}
The neural dynamics generating sensory, motor, and cognitive functions are commonly understood through field theories for neural population activity. 
Classic neural field theories are derived from highly simplified models of individual neurons, while biological neurons are highly complex cells.
Integrate-and-fire models retain a key nonlinear feature of neuronal activity: action potentials return the membrane potential to a nearly fixed reset value.
This nonlinear reset of the membrane voltage after a spike is absent from classic neural field theories.
Here, we develop a statistical field theory for networks of integrate-and-fire neurons with stochastic spike emission.
This reveals a new mean field theory for the activity in these networks, fluctuation corrections to the mean field dynamics, and a mapping to a self-consistent renewal process.
We use these to study the impact of the spike-driven reset of the membrane voltage on population activity.
The spike reset gives rise to a multiplicative, rate-dependent leak term in the mean field membrane voltage dynamics.
This leads to bistability between quiescent and active states in the mean field theory of homogenous and excitatory-inhibitory pulse-coupled networks. 
We uncover two types of fluctuation correction to the mean field theory, due to the nonlinear mapping from membrane voltage to spike emission and the nonlinear reset.
These can have competing effects, promoting and suppressing activity respectively. 
We then examine the roles of spike resets and recurrent inhibition in stabilizing network activity.
We calculate the phase diagram for inhibitory stabilization and find that an inhibition-stabilized regime occurs in wide regions of parameter space, consistent with
experimental reports of inhibitory stabilization in diverse brain regions.
Fluctuations narrow the region of inhibitory stabilization, consistent with their role in suppressing activity through spike resets.

\end{abstract}

\maketitle


\section{Introduction}
The activity of neuronal populations underlies sensory, motor, and cognitive functions. 
Mathematical theories for predicting the macroscopic activity of neural populations are a core tool of computational neuroscience, psychology, and psychiatry~\cite{freeman_mass_1975, coombes_large-scale_2010, bressloff_spatiotemporal_2011, moran_neural_2013}. 
These theories typically rely on neural activity equations, with variants also called rate equations, neural mass equations or, if placed on a spatial domain, neural field equations: 
\begin{equation} \label{eq:amari}
\partial_t {\bm v} = -{\bm v} + \bm{E} + {\bm J} \ast \phi(\bm{v}),  
\end{equation}
where bold terms denote a vector or matrix-valued function, $\phi$ is a single-unit nonlinearity applied elementwise, $\partial_t$ is the derivative with respect to time, and $\ast$ is a matrix convolution: $\bm{J} \ast \phi(\bm{v}) = \sum_j \int ds \; J_{ij}(s) \, \phi(v_j(t-s))$. 
These and similar equations are commonly understood as a coarse-grained model for large populations of neurons~\cite{grossberg_learning_1969, amari_characteristics_1971, amari_characteristics_1972, wilson_excitatory_1972, wilson_mathematical_1973}. 
Formally, they are a mean field theory for populations of neurons that switch between discrete active and quiescent states~\cite{ginzburg_theory_1994, ohira_master-equation_1993, buice_field-theoretic_2007, bressloff_stochastic_2010}, or for generalized linear point process models (Eq.~\ref{eq:linear_reset_mft}).

Biological neurons' membrane voltages, however, have complex nonlinear dynamics~\cite{hodgkin_quantitative_1952}. 
Neural field equations have been supplemented with some biophysical detail in an ad hoc fashion~\cite{bressloff_spatiotemporal_2011}. 
A principled mean field theory of more biophysical neuron models would expose how single-neuron biophysics shape macroscopic population activity~\cite{chow_before_2020}. 

Integrate-and-fire models, which replace the nonlinear dynamics of spike generation by a simple fire-and-reset rule for the membrane voltage, are fruitful tools for investigating how network structure and synaptic and neuronal biophysics shape macroscopic activity~\cite{brunel_single_2014, doiron_mechanics_2016}. 
The classic mean field theory of integrate-and-fire networks focuses on the density of membrane voltages across a population~\cite{knight_dynamics_1972}. 
If the net recurrent input to each neuron is a white Gaussian process, the membrane voltage density obeys a Fokker-Planck partial differential equation~\cite{ricciardi_diffusion_1977}. 
Numerical or special function solutions of that Fokker-Planck equation expose steady-state and weakly non-equilibrium population firing rates and pairwise statistics~\cite{amit_model_1997, brunel_dynamics_2000, lindner_transmission_2001, doiron_oscillatory_2004, lindner_theory_2005}.

The assumption of white Gaussian input currents is, however, inconsistent with the resulting temporally colored spike train statistics~\cite{lindner_superposition_2006}. 
In some cases, the Fokker-Planck approach for the population voltage density can be extended to temporally structured fluctuations~\cite{moreno-bote_auto-_2006, schwalger_statistical_2015, vellmer_theory_2019}. 
Alternatively, for generalized integrate-and-fire neurons with stochastic spike emission, population firing rates and pairwise statistics can be predicted from the density of inter-spike times rather than the density of membrane voltages~\cite{gerstner_time_1995, gerstner_population_2000, meyer_temporal_2002, deger_fluctuations_2014, dumont_stochastic-field_2017, schwalger_towards_2017}. 
Population density approaches expose approximate low-dimensional dynamics through eigenfunctions of the density evolution operator~\cite{mattia_population_2002, pietras_low-dimensional_2020}. 

Here, we study the integrate-and-fire model with stochastic spike emission. 
We construct the full joint probability density functional of a neuronal network's spike trains and membrane voltages using the response variable path integral formalism~\cite{martin_statistical_1973, dominicis_techniques_1976, janssen_lagrangean_1976, jensen_functional_1981}. 
This formalism is commonly applied to non-spiking models, where it has exposed chaotic and metastable regimes~\cite{sompolinsky_chaos_1988, stern_dynamics_2014, bressloff_path-integral_2015, van_meegen_large-deviation_2021}, the memory capacity of recurrent networks~\cite{schuecker_optimal_2018, pereira_attractor_2018}, and the dynamical impact of computationally or biologically constrained connectivity~\cite{aljadeff_transition_2015, mastrogiuseppe_linking_2018, schuessler_dynamics_2020, landau_macroscopic_2021}. 
It has also been applied to spiking models without spike resets or in a phase formulation~\cite{buice_dynamic_2013, kadmon_transition_2015, harish_asynchronous_2015, mastrogiuseppe_intrinsically-generated_2017, ocker_linking_2017, brinkman_predicting_2018, todorov_stability_2019, kordovan_spike_2020}.	

The joint density functional exposes a new simple, deterministic mean field theory for stochastic integrate-and-fire networks: activity equations like Eq.~\ref{eq:amari} with an additional rate-dependent leak. 
This novel nonlinearity qualitatively shapes networks' macroscopic dynamics. 
We study networks in an increasing order of complexity, progressing from uncoupled neurons to single-population recurrent networks and then networks with multiple cell types. 
Spike resets can stabilize strongly coupled excitatory networks with unbounded spike intensity functions. 
We uncover bistable regimes in homogenous and excitatory-inhibitory networks. 
Examining the impact of fluctuations on the activity using renewal theory and a self-consistent Gaussian approximation with colored noise, we find that due to the nonlinearity of the spike reset, fluctuations suppress activity.  

In the classic neural activity equations, inhibitory feedback is necessary to stabilize strong recurrent excitation~\cite{griffith_stability_1963, wilson_excitatory_1972}. 
A paradoxical reduction of inhibitory activity after inhibitory stimulation is a signature of an inhibition-stabilized regime~\cite{tsodyks_paradoxical_1997} and is observed in diverse mammalian cortices~\cite{ozeki_inhibitory_2009, kato_network-level_2017, adesnik_synaptic_2017, sanzeni_inhibition_2020}. 
We find that the phase diagram for excitatory-inhibitory networks includes wide regions of paradoxical responses, suggesting a generic mechanism for their widespread experimental observation.
Spiking fluctuations narrow the region of inhibitory stabilization, consistent with their intrinsically stabilizing effect through resets of the membrane voltage. 




\section{Stochastic integrate-and-fire model}
We introduce the stochastic leaky integrate-and-fire (LIF) model in discrete time first and then take a continuous-time limit.
At each small time step $t \in [T]$, of width $dt$, neuron $i \in [N]$ generates $dn_{it} \in \{0, 1\}$ spikes. ($n_{it}$ is the cumulative spike count of neuron $i$ at time $t$.) 
Neuron $i$ receives inputs $d\bm{n}$ through weighted synaptic filters $\bm{J}$. 
It also has a resting voltage $E_i$, which may depend on external applied currents. 
We take $dn_{it}$ to be generated as a Bernoulli random variable with spike probability $f(v_{it}) \, dt$, for some intensity function $0 \leq f(v) \leq dt^{-1}$. 
After a spike is emitted, that neuron's membrane voltage is reset to within $\mathcal{O}(dt)$ of the reset value $r$.
If $f(v) = \theta(v-b) / dt$, where $\theta(x)$ is the Heaviside step function, the deterministic LIF neuron with threshold $b$ is recovered~\cite{zhou_investigating_2021}. 
In the continuous-time limit (Appendix~\ref{app:density}),
\begin{equation} \label{eq:lif}
\partial_t \bm{v}(t) = \frac{1}{\tau}\left(-{\bm v}(t) + {\bm E}(t) + ({\bm J} \ast \dot{\bm n})(t)\right) - \dot{\bm n}(t^+) \left({\bm v}(t)-r\right).
\end{equation}
Here, $\dot{n}_i(t^+) \equiv \partial_t n(t^+)$, and $t^+ = t + \epsilon$ for an infinitesimal $\epsilon > 0$ so the membrane potential is reset at time $t$ by an immediately preceding spike.
Each $\dot{n}_i(t)$ is an inhomogenous Poisson process with intensity $f(v_i(t))$. The Poisson spike emission arises as the continuous-time limit of the discrete-time Bernoulli spike train.
The last term in Eq.~\ref{eq:lif} is the reset of the membrane voltage after a spike. 
This nonlinear coupling between the spike train and membrane voltage is the key feature of this model compared to generalized linear models.  
(See Appendix~\ref{app:absolute_refractory} for a discussion of absolute refractory periods in this model.)
We will non-dimensionalize the model, measuring time relative to $\tau$ and shifting ${\bm v}$ and $E$ by the reset $r$. 

Eq.~\ref{eq:lif} is a set of coupled stochastic differential equations with multiplicative Poisson noise. 
The expected trajectory obeys
\begin{equation} \begin{aligned} \label{eq:v_avg}
\partial_t \langle \bm{v} \rangle =& -\langle \bm{v} \rangle + {\bm E} + {\bm J} \ast \langle \dot{\bm n} \rangle - \langle \dot{\bm n} \rangle \langle \bm{v} \rangle - \llrrangle{\dot{\bm n} \bm{v}}, \\
\langle \dot{\bm n} \rangle =& \langle f( \bm{v}) \rangle.
\end{aligned} \end{equation}
where $\langle \rangle$ denotes a moment and $\llrrangle{}$ denotes a cumulant. (Here we suppress the explicit time dependencies, as well as the infinitesimal time shift in the reset term. Moving forwards, we will often continue to suppress those.)
To compute those requires the joint density functional of the membrane voltages and spike trains. 
In the response variable path integral formalism, it is (Appendix \ref{app:density}):
\begin{equation} \begin{aligned} \label{eq:lif_action}
p[\bm{v}, \dot{\bm{n}}] =&  \int \mathcal{D}\tilde{\bm{v}} \int \mathcal{D} \tilde{\bm{n}} \; \exp \left(-S[\bm{v}, \dot{\bm n}, \tilde{\bm{v}}, \tilde{\bm{n}}] \right), \\
S[\bm{v}, \dot{\bm{n}}, \tilde{\bm{v}}, \tilde{\bm{n}}] =&  \tilde{\bm v}^T \left(\partial_t \bm{v} +{ \bm v} - {\bm E} -{\bm J} \ast \dot{\bm n} + \dot{\bm n} {\bm v} \right)\\
&+ \tilde{\bm n}^T \dot{\bm n} - \left( \exp (\tilde{\bm n})-1 \right)^T {\bm f}.
\end{aligned} \end{equation}
Here, ${\bm x}^T {\bm y} = \sum_i \int dt\, x_i(t) \, y_i(t)$ is the functional inner product and $f_i(t) = f(v_i(t))$. $S$ is the action functional. $\tilde{\bm{n}}, \tilde{\bm{v}}$ are purely imaginary auxiliary variables, called the response variables because joint moments with them measure responses to fluctuations in the activity. 

This density has the $N$-dimensional deterministic mean field theory
\begin{equation} \label{eq:lif_mft}
\partial_t \bar{\bm{v}} = - \bar{\bm{v}} + {\bm E} + {\bm J} \ast \bar{{\bm f}} - \bar{\bm{f}} \bar{ \bm{v}}.
\end{equation} 
The mean field value of $\dot{n}_i$ is $\bar{f}_i = f(\bar{v}_i)$. The $N$-dimensional mean field theory is an approximation of Eq.~\ref{eq:v_avg}.
If we assume that
\begin{enumerate}
\item the expectation of the spike trains is $\langle \dot{\bm n} \rangle = f(\langle \bm{v} \rangle )$ 
\item the spikes and membrane voltage are independent so that the joint cumulant $\llrrangle{ \dot{\bm n} {\bm v} } = 0$,
\end{enumerate}
then Eq.~\ref{eq:v_avg} reduces to Eq.~\ref{eq:lif_mft}. 
Assumption (1) is only correct if $f$ is linear.
Assumption (2) is generally incorrect, although it may be a good approximation if $\langle {\bm v} \rangle \langle \dot{\bm n}\rangle \gg \llrrangle{ \dot{\bm n} {\bm v} }$.

Formally, we expand the configuration variables $\bm{v}, \dot{\bm n}$ around their mean values. 
The mean field theory is then the result of a saddle point approximation for integrals over the fluctuations (Appendix~\ref{app:effective_action}). 
This corresponds to assuming fluctuations are negligible so $p[\dot{\bm{n}}]  =  \delta [\dot{\bm{n}} - f(\bm{v})]$, or equivalently, truncating the action at linear order in $\tilde{\bm n}$.
This implies assumptions (1) and (2), so Eq.~\ref{eq:v_avg} reduces to Eq.~\ref{eq:lif_mft}. 
This approach also exposes fluctuation corrections to the mean field theory.
We will see in the next section that the two nonlinearities in Eq.~\ref{eq:lif} impart different fluctuation corrections to the mean field theory.
First, we compare the mean field theory, Eq.~\ref{eq:lif_mft}, to the classic activity equations, Eq.~\ref{eq:amari}.


The mean field dynamics of Eq.~\ref{eq:lif_mft} differ from Eq.~\ref{eq:amari} in two ways. 
The first is the presence of the reset term $-f(v_i) \, v_i $. 
The second is in the interpretation of the nonlinearity $f$.
Here, $f$ determines the instantaneous spike emission probability as a function of the membrane voltage and is typically required to be non-saturating so that the neuron is guaranteed to spike if $v_i \rightarrow \infty$. (This is not mathematically necessary; $f$ could be chosen to saturate at a finite value. In discrete time, $f$ must be bounded by $1/dt$ so the spike probability does not exceed 1.)
In the microscopic binary switching model underlying Eq.~\ref{eq:amari}, the nonlinearity $\phi$ determines the single-neuron transition rates from quiescence to activity and is typically chosen as a sigmoid to prevent unbounded activity. 
In either case, the nonlinearity $f$ or $\phi$ is a property of individual neurons.

Can we map the new mean field theory, Eq.~\ref{eq:lif_mft}, onto the classic activity equations, Eq.~\ref{eq:amari}, with an effective nonlinearity $\phi$ that includes the effect of the rate-dependent leak? 
Requiring $\bm{J} \bm{\phi} = \bm{J} \bm{f} - \bm{v} \bm{f} $, with $(\bm{v} \bm{f})_i = v_i f(v_i)$, we find that if the coupling $\bm{J}$ has a left inverse,
\begin{equation}
\bm{\phi} (\bm{v}) = \bm{f} - \bm{J}^{-1} (\bm{v} \bm{f}) .
\end{equation}
So to map the mean field theory of Eq.~\ref{eq:lif_mft} onto the classic activity equations, the effective nonlinearity $\phi$ depends explicitly on the coupling $\bm{J}$; it is no longer a single-neuron nonlinearity. 
If there are linear self-interactions and inter-neuronal coupling is weak so that $\bm{J}$ is diagonally dominant, the effective nonlinearity will be approximately a single-neuron property.

The other classic form of rate equation is $\tau \, \partial_t {\bm v} = -{\bm v} + \phi({\bm J} \ast {\bm v} + {\bm E})$. 
This is also a mean field theory of binary switching neurons~\citep{ohira_master-equation_1993, buice_field-theoretic_2007, bressloff_stochastic_2010}.
Here, $v$ is commonly understood as a mean field description of the firing rate or proportion of active neurons in a population, rather than the membrane potential or synaptic drive~\cite{wilson_excitatory_1972, wilson_mathematical_1973}. 
The two types of activity equation differ in their assumptions about the dominant synaptic or neuronal timescales~\cite{pinto_quantitative_1996, bressloff_stochastic_2010}.

To map Eq.~\ref{eq:lif_mft} onto this would require $ \phi((\bm{J} \ast {\bm v})_i + E_i) = - v_i f(v_i) + \sum_j J_{ij} \ast f(v_j) + E_i$. 
In general, mapping Eq.~\ref{eq:lif_mft} onto this may require the nonlinearity to be a function of the coupling operator, activity variable, and baseline drive separately, rather than a function of their sum. 

Mapping Eq.~\ref{eq:lif_mft} onto the classic activity equations can thus introduce nonlinearities tailored to a specific LIF network, rather than as single-neuron input-rate functions. 
This mapping is, however, not necessary.
The mean field dynamics of Eq.~\ref{eq:lif_mft} are amenable to direct analysis. 

\section{Impact of spike reset and fluctuations on single-neuron activity} \label{sec:reset}
\begin{figure} \includegraphics[]{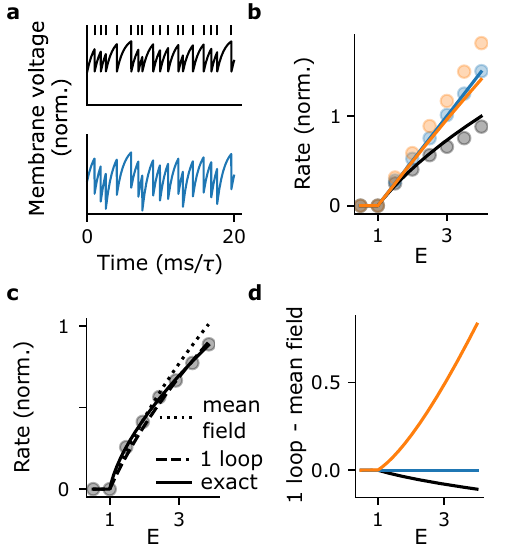}
\caption{Impact of fluctuations on firing rates through spike resets and nonlinear intensity functions. {\bf a)} Membrane voltage traces of the stochastic LIF neuron (top, black) and a neuron with linear resets (bottom, blue). For comparison, the two neurons are forced to have the same spike times (in this panel only). {\bf b)} Firing rate vs resting voltage, $E$, for three models. Black: the stochastic LIF neuron with a threshold-linear intensity function, $f(v) = \lfloor v-1 \rfloor_+$. Blue: the linear-reset model with a matched intensity function, $f_L(v) = f(v)$. Orange: the linear-reset model with matched mean field membrane voltages, $f_M(v) = vf(v)$. Dots: simulation. Solid curves: mean field predictions (Eqs.~\ref{eq:lif_fI_mft},~\ref{eq:linear_reset_fI_mft},~\ref{eq:linear_match_fI_mft}). {\bf{c})} Impact of fluctuations on the stochastic LIF neuron's firing rate. Dotted: mean field prediction. Dashed: self-consistent one-loop prediction, accounting for Gaussian fluctuations around the expected voltage and rate (Eq.~\ref{eq:lif_feynman_1loop}). Solid: exact renewal theory prediction from Eq.~\ref{eq:lif_mfpt_exact}. {\bf d)} Difference between the one-loop and mean-field rates for the three models.
\label{fig:intro} }
\end{figure}

We now examine the steady-state input-rate transfer of a single neuron or, equivalently, an uncoupled population. 
The mean field firing rate, $\bar{f}$, is given by equilibria of Eq.~\ref{eq:lif_mft} with $\bm{J}={\bf 0}$ and constant $E$. 
We consider neurons with threshold-power law spike probability functions, $f(v)=\lfloor v -1 \rfloor_+^a$, which match the effective nonlinearity of mechanistic spiking models and biological neurons in fluctuation-driven regimes~\cite{miller_kenneth_d._neural_2002, hansel_how_2002, priebe_contribution_2004, priebe_mechanisms_2006, linaro_correlation_2019}. (The membrane voltage has been non-dimensionalized to set the threshold for spike generation at $v=1$.)
For simplicity, we take a threshold-linear neuron with $f(v)=\lfloor v - 1 \rfloor_+$ so the equilibrium solution to the mean field equation is
\begin{equation} \begin{aligned} \label{eq:lif_fI_mft}
\bar{f} = \lfloor \sqrt{E}-1\rfloor_+.
\end{aligned} \end{equation}
The mean field theory for the stochastic LIF neuron predicts its equilibrium firing rate as a function of its membrane voltage (Fig. \ref{fig:intro}b, black line vs dots). 
At higher rates, Eq.~\ref{eq:lif_fI_mft} overpredicts the true firing rates. Since the mean field theory neglects all fluctuations, fluctuations suppress activity in the stochastic LIF model. 

For comparison, consider a stochastic LIF model with a linear reset: each spike causes a decrease in the membrane voltage of size $r$ (Fig.~\ref{fig:intro}a, blue)~\cite{gerstner_universality_1992}. The action for that model is
\begin{equation} \begin{aligned} \label{eq:linear_reset_action}
S[\bm{v}, \dot{\bm{n}}, \tilde{\bm{v}}, \tilde{\bm{n}}] =&  \tilde{\bm v}^T \left(\partial_t \bm{v} +{ \bm v} - {\bm E} -{\bm J} \ast \dot{\bm n} + \dot{\bm n}r\right) \\
&+ \tilde{\bm n}^T \dot{\bm n} - \left( \exp (\tilde{\bm n})-1 \right)^T {\bm f}_L
\end{aligned} \end{equation}
with the $N$-dimensional mean field theory
\begin{equation} \label{eq:linear_reset_mft}
\partial_t \bar{\bm v} = - \bar{\bm v} + {\bm E} + {\bm J} \ast \bar{\bm f} - \bar{\bm f}r .
\end{equation}
This has a similar form to the classic activity equation, Eq.~\ref{eq:amari}, and can be directly mapped onto it with the substitution $J_{ii}(s) \rightarrow J_{ii}(s) - r\delta(s)$. For this reason, we say that Eq.~\ref{eq:amari} is a mean field theory for a stochastic LIF neuron with linear resets, which is an example of a generalized linear model or 0th order spike response model~\cite{gerstner_neuronal_2014}. 
The mean field firing rate of the uncoupled linear-reset model, with $r=v_{th}=1$, is
\begin{equation} \label{eq:linear_reset_fI_mft}
\bar{f}_L = \left \lfloor \frac{E-1}{2} \right \rfloor_+ .
\end{equation}
For a peri-threshold stimulus, $E = 1+\epsilon$ in Eqs.~\ref{eq:lif_fI_mft} and~\ref{eq:linear_reset_fI_mft}, $\bar{f} = \epsilon/2 + \mathcal{O}(\epsilon^2) \approx \bar{f}_L$ and the mean field theories of the stochastic LIF and linear-reset models match for infinitesimal firing rates. At finite rates, however,  the linear-reset model provides a poor prediction for the stochastic LIF neuron (Fig. \ref{fig:intro}b, blue vs black). 

Instead of matching the intensity functions of the two models, we could match their mean-field membrane voltage by giving the linear-reset model the intensity function $f_M(v) = v f(v)$. ($f(v)$ is the intensity function of the stochastic LIF neuron.) The mean-field rate of this matched linear-reset model is
\begin{equation} \label{eq:linear_match_fI_mft}
\bar{f}_M = \sqrt{E} \lfloor \sqrt{E}-1\rfloor_+.
\end{equation}
For the matched linear-reset model, the mean field firing rate underpredicts the true activity level (Fig.~\ref{fig:intro}b, orange line vs dots), so fluctuations promote activity. Why do fluctuations suppress activity in the stochastic LIF model but promote activity in the matched linear-reset model?

In $\partial_t \langle \bm{v} \rangle$, we need to account for (1) the nonlinearity in the intensity function and (2) the nonlinear spike reset.
To that end, we expand the membrane voltage and spike trains around their means to derive an expansion for the action that self-consistently accounts for the impact of fluctuations (the loop expansion of the effective action; Appendix~\ref{app:effective_action}).
This allows us to derive diagrammatic corrections to the mean field theory.
Loop diagrams measure the influence of higher-order activity statistics on lower-order statistics. 
There are loop corrections to the mean field theory when the model has a nonlinearity.

The stochastic LIF has two nonlinearities: the intensity function and the nonlinear spike reset. These give rise to fluctuation corrections in the mean voltage and rate:
\begin{equation} \begin{aligned} \label{eq:lif_feynman_1loop}
0 =& \partial_t \bar{v} + \bar{v} + \bar{v} \bar{n} - E +
 \vcenter{ \hbox{
 \feynmandiagram[scale=0.5][horizontal=a to b, inline=(b.base)]{
a[empty dot] --[half left, photon] b[dot] --[half left] a,
 };  }}  \\
 0 =&\bar{n} - f(\bar{v}) -
\vcenter{ \hbox{
 \feynmandiagram[scale=0.5][ inline=(b.base), horizontal=a to b, small]{
a[empty dot] --[half left, photon] b[dot] --[half left, photon] a,
 }; }} .
\end{aligned} \end{equation} 
Without the one-loop diagrams, these reduce to the mean field theory of Eq.~\ref{eq:lif_mft} with $\bm{J}=0$. The one-loop diagrams measure the impact of two-point fluctuations on the mean through the two nonlinearities of the intensity function and spike reset. 
%

In field theoretic terms, these loop diagrams represent proper vertex corrections to the effective action (Appendix~\ref{app:effective_action}). We can also understand them by comparing Eq.~\ref{eq:v_avg} and Eq.~\ref{eq:lif_feynman_1loop}. In their first lines, the loop diagram $\vcenter{ \hbox{
 \feynmandiagram[scale=0.5][horizontal=a to b, inline=(b.base)]{
a[empty dot] --[half left, photon] b[dot] --[half left] a,
 };  }}$ is an approximation of $\llrrangle{\dot{n} v}$. In the second line of Eq.~\ref{eq:lif_feynman_1loop} , the loop diagram originates in a Taylor expansion of $f(v)$ around $f(\langle v \rangle)$ in Eq.~\ref{eq:v_avg}; the loop diagram in the second line approximates $\frac{f^{(2)}}{2} \llrrangle{v^2}$. We next discuss these approximations.



The edges in the Feynman diagrams correspond to factors of the linear response of the configuration to a fluctuation, $\bar{\bm{\Delta}}$, also called a propagator.
Since the model has two configuration variables, each with a corresponding response variable, there are four types of propagator: 
\begin{enumerate}
\item the spike response to a spike fluctuation $\bar{\Delta}_{ n, \tilde{n}}$, 
\item the voltage response to a spike fluctuation $\bar{\Delta}_{ v, \tilde{n}}$,
\item the spike response to a voltage fluctuation $\bar{\Delta}_{ n, \tilde{v}} $, and 
\item the voltage response to a voltage fluctuation $ \bar{\Delta}_{ v, \tilde{v}}$. 
\end{enumerate}
We represent them with the edges
\begin{equation} \begin{aligned} \label{eq:feynman_propagators}
 \bar{\Delta}_{ n, \tilde{n}} =&
 \vcenter{ \hbox{
 \feynmandiagram[horizontal=a to b, inline=(a.base)]{
a --[] b,
 }; }}\\
\bar{\Delta}_{ v, \tilde{n}} =&
\vcenter{ \hbox{
\feynmandiagram[horizontal=a to b, inline=(a.base)]{
a --[photon] b [],
 }; }} \\
\bar{\Delta}_{ n, \tilde{v}} =&
\vcenter{ \hbox{
\feynmandiagram[horizontal=a to b, inline=(a.base)]{
a --[gluon] b [],
 }; }} \\
 \bar{\Delta}_{ v, \tilde{v}} =&
 \vcenter{ \hbox{
\feynmandiagram[horizontal=a to b, inline=(a.base)]{
a --[scalar] b [],
 }; }}
 \end{aligned} \end{equation}
Only the first two of these edges appear in the one-loop equations of motion for the mean voltage and rate. 
The vertex $\feynmandiagram{a[dot]};$ represents the intensity, $f(\bar{v})$. 
Each diagram also has a vertex $\feynmandiagram{a[empty dot]};$. 
These vertices have different origins in the two diagrams, corresponding to either the spike reset or intensity function. (The two types of $\feynmandiagram{a[empty dot]};$ vertex can be distinguished by their incoming edges.)
The definition of the linear response functions for the stochastic LIF model are given in Appendix~\ref{app:feynman_rules}, along with the Feynman rules for perturbative corrections to the mean field theory (see also Appendix~\ref{app:effective_action}).

In Eq.~\ref{eq:v_avg}, at a stationary state
\begin{equation} \begin{aligned} \label{eq:bubble_reset}
\llrrangle{\dot{n}(t^+) \, v(t)} \approx &
\vcenter{ \hbox{
 \feynmandiagram[scale=0.5][horizontal=a to b, inline=(b.base)]{
a[empty dot] --[half left, photon] b[dot] --[half left] a,
 };  }} \\
 =& \lim_{t^+ \rightarrow t} \int \frac{d \omega}{2 \pi} \; e^{i \omega \left(t^+ - t\right)} \, \bar{\Delta}_{ n, \tilde{n}}(\omega) \, \bar{\Delta}_{ v, \tilde{n}}(-\omega) \, f(\bar{v}) \\
 =& \frac{f^{(1)} \bar{v}^2 }{2 \left(1 + \bar{n} + f^{(1)} \bar{v} \right)} f(\bar{v}).
\end{aligned} \end{equation}
We have written the cumulant for a single neuron, dropping the neuron index implicit in Eq.~\ref{eq:v_avg} and leveraged the stationarity assumption so that $\bar{\Delta}(t, t') = \bar{\Delta}(t'-t) = \mathcal{F}^{-1}[\bar{\Delta}(\omega)]$, with the Fourier transform convention $\mathcal{F}[\bar{\Delta}(s)] = \int ds \, \Delta(s) e^{-i \omega s}$. The limit $t^+ \rightarrow t$ is taken from the right, $t^+ > t$. $f^{(p)}$ is the order-$p$ derivative of the intensity function $f$, evaluated at $\bar{v}$. This diagram, and the vertex $\feynmandiagram{a[empty dot]};$ in it, arise from the spike reset term of Eq.~\ref{eq:lif}.
If the neuron is in an active steady state, $\bar{v} > 1$ and $f(\bar{v}), \bar{n} > 0$. If $f^{(1)} > 0$, every term in this approximate cumulant is positive. It thus decreases $\partial_t \bar{v}$ in Eq.~\ref{eq:lif_feynman_1loop}.
 
The equation of motion governing $\bar{n}$ also has a loop correction. In the second line of Eq.~\ref{eq:v_avg}, we expand $f(v)$ around the mean. Truncating at second order, in a stationary state
\begin{equation} \begin{aligned}  \label{eq:bubble_hazard}
\frac{f^{(2)}}{2} \llrrangle{v(t) \, v(t)} \approx& 
\vcenter{ \hbox{
 \feynmandiagram[scale=0.5][ inline=(b.base), horizontal=a to b, small]{
a[empty dot] --[half left, photon] b[dot] --[half left, photon] a,
 }; }}  \\
 =& \frac{f^{(2)}}{2}  \int \frac{d \omega}{2 \pi} \; \bar{\Delta}_{ v, \tilde{n}}(\omega) \, \bar{\Delta}_{ v, \tilde{n}}(-\omega) \, f(\bar{v}) \\
 =& \frac{f^{(2)}}{2} \frac{\bar{v}^2}{2 \left(1 + \bar{n} + f^{(1)} \bar{v} \right)} f(\bar{v}).
\end{aligned} \end{equation}
This diagram exists in both the stochastic LIF and linear-reset models. It arises from the nonlinear intensity function, $f$, and vanishes almost everywhere if $f$ is threshold-linear. The vertex $\feynmandiagram{a[empty dot]};$ in it carries the factor of $f^{(2)} / 2$.
This contribution impacts the mapping from membrane potential to firing rate. Similarly to above, the approximate cumulant $\llrrangle{v(t) \, v(t)}$ is positive in an active state with non-decreasing $f$.
So if $f^{(2)} > 0$ fluctuations will promote activity and vice versa. The curvature also determines the magnitude of this contribution.

Evaluating the one-loop predictions at an equilibrium of $\bar{v}$, the stochastic LIF with a threshold-linear intensity function has the one-loop equilibrium
\begin{equation} \begin{aligned} \label{eq:lif_single_neuron_1loop}
\bar{v} =& \frac{1}{10}\left(1+ \sqrt{1+80E} \right) \\
\bar{n} =& \lfloor \bar{v} - 1\rfloor_+
\end{aligned} \end{equation} 
Comparing the one-loop and mean field predictions, we see that the negative vertex factor leads to a suppression of activity (Fig.~\ref{fig:intro}c, dotted vs dashed; Fig.~\ref{fig:intro}d, black). This occurs because the nonlinear spike reset negatively couples the mean membrane voltage to joint fluctuations in the spikes and membrane voltage. 

We can also calculate the rate of the threshold-linear stochastic LIF neuron exactly. Due to the nonlinear reset mechanism, the spike train is a renewal process. Standard results of renewal theory expose its rate~\cite{cox_point_1980}.
With a constant drive $E$, the membrane voltage evolves after a spike at time $t$ as
$v(t+s) = E \left(1 - \exp(-s) \right)$,
with $v(t) = 0$. The time-averaged firing rate is the inverse of the mean interspike interval: $\langle \dot{n} \rangle = 1 / \langle s \rangle$.
For threshold-linear $f$, the mean interspike interval is
\begin{equation} \begin{aligned} \label{eq:lif_mfpt_exact}
\langle s \rangle 
=& \ln \left( \frac{E}{E-1} \right) + \left(\frac{E-1}{e} \right)^{1-E} \gamma(E-1, E-1) .
\end{aligned} \end{equation} 
$\gamma(x, y)$ is the lower incomplete gamma function. The term $\ln \left( \frac{E}{E-1} \right)$ is the time for $v(t)$ to reach the threshold value of 1; the second term is the mean first spike time after that. 
The one-loop prediction matches the true f-I curve better than the mean-field theory (Fig.~\ref{fig:intro}c). 

In the uncoupled linear-reset model, the one-loop equations of motion are
\begin{equation} \begin{aligned} \label{eq:feynman_1loop_linear_reset}
0 =& \partial_t \bar{v} + \bar{v} + r\bar{n} - E \\
 0 =&\bar{n} - f(\bar{v}) -
\vcenter{ \hbox{
 \feynmandiagram[scale=0.5][ inline=(b.base), horizontal=a to b, small]{
a[empty dot] --[half left, photon] b[dot] --[half left, photon] a,
 }; }} .
\end{aligned} \end{equation}
The linear-reset model has the same four types of propagator as the stochastic LIF, although their definitions differ between the two models due to the different spike reset mechanisms (Appendix~\ref{app:linear_reset}). 
The matched linear-reset model ($f_M(v) = v \lfloor v-1\rfloor_+$, with $r=1$) has the one-loop equilibrium
\begin{equation} \begin{aligned}
\bar{v} = \frac{1}{8} \left(-1 + \sqrt{64E +17} \right) \\
\bar{n} = \left(\bar{v} + \frac{1}{4}\right) \lfloor \bar{v}-1 \rfloor_+ 
\end{aligned} \end{equation} 
The term $\frac{1}{4} \lfloor \bar{v}-1 \rfloor_+$ is the result of the loop correction to the firing rate, leading to an increase in the rate compared to the mean-field rate. The expected membrane potential still reaches the threshold value $\bar{v}=1$ at $E=1$, and the one-loop membrane voltage is greater than the mean-field membrane voltage $\sqrt{E}$. Close to threshold ($E=1+\epsilon$), the one-loop voltage is $\bar{v} = 1 + 4 \epsilon / 9 + \mathcal{O}(\epsilon^2)$ while for the mean-field theory, $\bar{v} = 1 + \epsilon/2 + \mathcal{O}(\epsilon^2)$. Similarly, for large $E$ the one-loop membrane potential is approximately $1 + \sqrt{E}$.

In summary: fluctuations suppress activity in the stochastic LIF neuron because the nonlinear spike reset negatively couples the mean membrane voltage to joint spike-voltage fluctuations. 
In the linear-reset model, the only nonlinearity arises from the intensity function. 
To match the mean-field voltage of the stochastic LIF neuron, the linear neuron's intensity function is $f_M(v) = vf(v)$, where $f(v)$ is the stochastic LIF intensity function. 
Since $f(v)$ had non-negative curvature, $f_M$ has positive curvature. 
So, fluctuations of the membrane voltage promote spiking activity in the matched linear-reset model. 
A stochastic LIF network with a nonlinear intensity function may have contributions from both diagrams, so that the two nonlinearities compete to determine whether fluctuations suppress or promote activity.

\section{Homogenous networks}
Biological neural networks are coupled. 
We will seek a low-dimensional description of the population activity that accounts for synaptic coupling.
Here, we study the simplest case: networks where the connectivity between neurons is homogenous, so we take the synaptic weights between neurons from a distribution with negligible second- and higher-order cumulants. 
We assume that the mean synaptic weight is $\mathcal{O}(1/N)$ so the total synaptic weight onto a neuron is $\mathcal{O}(1)$. 
An exemplar of this case is a network with weak ($J_{ij} \sim 1/N$) but potentially dense (connection probability $\sim 1$) connections (Appendix~\ref{app:mft_homog}). 
To examine the interaction between synaptic connectivity, subthreshold dynamics, and stochastic spike emission in shaping network activity, we will average the partition functional for the activity (equivalently, average the moment generating functional) over realizations of the synaptic connectivity (Appendix \ref{app:mft_homog}). 
In the limit of large $N$, the density factorizes over the neurons, yielding the partition functional
\begin{equation} \begin{aligned} \label{eq:saddle_point_Z_1pop_homog}
Z^* =& \int \mathcal{D} v  \int \mathcal{D} \dot{n}  \int \mathcal{D} \tilde{v}  \int \mathcal{D} \tilde{n} \; \exp \bigg( \\
& -\tilde{v}^T \left( \partial_t v + v - E - J \ast \langle \dot{n} \rangle + \dot{n} v \right) \\
&- \tilde{n}^T \dot{n} + \left(e^{\tilde{n}}-1 \right)^T f \bigg).
\end{aligned} \end{equation}
The result is a population of independent stochastic LIF neurons, each receiving a self-consistent mean field input $J \ast \langle \dot{n} \rangle$, where $\langle \dot{n} \rangle$ is the population-averaged spike train. For self-averaging connectivity, the result describes the typical behavior of an individual network and the population average $\langle \dot{n} \rangle$ matches the ensemble averaged rate.
Since the density factorizes, we drop the neuron index. 
Robert \& Touboul proved convergence to these mean field dynamics~\cite{robert_dynamics_2016}. 
The connectivity has been reduced to its mean, $J$, which would be equivalent to assuming a network with all-to-all connectivity. $J$ can be either positive or negative. If the connectivity had non-negligible higher cumulants, these would give rise to corresponding fluctuations in the membrane potential (Appendix~\ref{app:mft_homog}).
This population-averaged mean field theory is one-dimensional not because the neurons are synchronized, but because they spike independently given a self-consistent mean field input.

If the network is in an asynchronous state so $\langle \dot{n} \rangle$ is constant in time, after a spike at time $t$ the membrane voltage obeys
\begin{equation} \label{eq:lif_postspike}
v(t+s) = \left(E + J \langle \dot{n}\rangle \right) \left(1 - \exp(-s) \right)
\end{equation}
and the spike train is a renewal process. 
(We write $J$ for the integral of the coupling kernel $J(s)$.) 
With a threshold-linear intensity function, the mean interspike interval is
\begin{equation} \begin{aligned} \label{eq:lif_1pop_exact}
\langle s \rangle 
=& \ln \left( \frac{C}{C-1} \right) + \left(\frac{C-1}{e} \right)^{1-C} \gamma(C-1, C-1),
\end{aligned} \end{equation} 
where $C = E + J \langle \dot{n} \rangle$. In a stationary state, the rate is the inverse of the mean interspike interval: $\langle \dot{n} \rangle = 1 / \langle s \rangle$, which allows us to find self-consistent solutions of Eq.~\ref{eq:lif_1pop_exact} numerically.

The mean field (tree-level) equation of motion for the membrane voltage is 
\begin{equation} \label{eq:lif_mft_1pop}
0 = \partial_t \bar{v} + \bar{v} + \bar{v}f(\bar{v}) - E - J \ast f(\bar{v}),
\end{equation}
with $f(\bar{v})$ the mean field approximation of $\dot{n}$.
As in the $N$-dimensional mean field theory of Eq.~\ref{eq:lif_mft}, this neglects all fluctuations, so we expect that it will not be quantitatively correct. Since the spike trains are conditionally Poisson, those fluctuations are driven by the expected intensity. We thus expect that Eq.~\ref{eq:lif_mft_1pop} should be a good approximation when the true firing rate is low. As we will see below, it can provide a good qualitative description of the population dynamics, including bifurcations from quiescence.
The leading-order description of fluctuations is given by the one-loop equations of motion,
\begin{equation} \begin{aligned} \label{eq:feynman_1loop_1pop}
0 =& \partial_t \bar{v} + \bar{v} + \bar{v} \bar{n} - E - J \ast \bar{n} +
 \vcenter{ \hbox{
 \feynmandiagram[scale=0.5][horizontal=a to b, inline=(b.base)]{
a[empty dot] --[half left, photon] b[dot] --[half left] a,
 };  }} \\
 0 =&\bar{n} - f(\bar{v}) -
\vcenter{ \hbox{
 \feynmandiagram[scale=0.5][ inline=(b.base), horizontal=a to b, small]{
a[empty dot] --[half left, photon] b[dot] --[half left, photon] a,
 }; }} .
\end{aligned} \end{equation}
The one-loop contributions are given by Eqs.~\ref{eq:bubble_reset},~\ref{eq:bubble_hazard}.

\section{Bistable activity in homogenous networks}
\begin{figure*}[ht!] \includegraphics[]{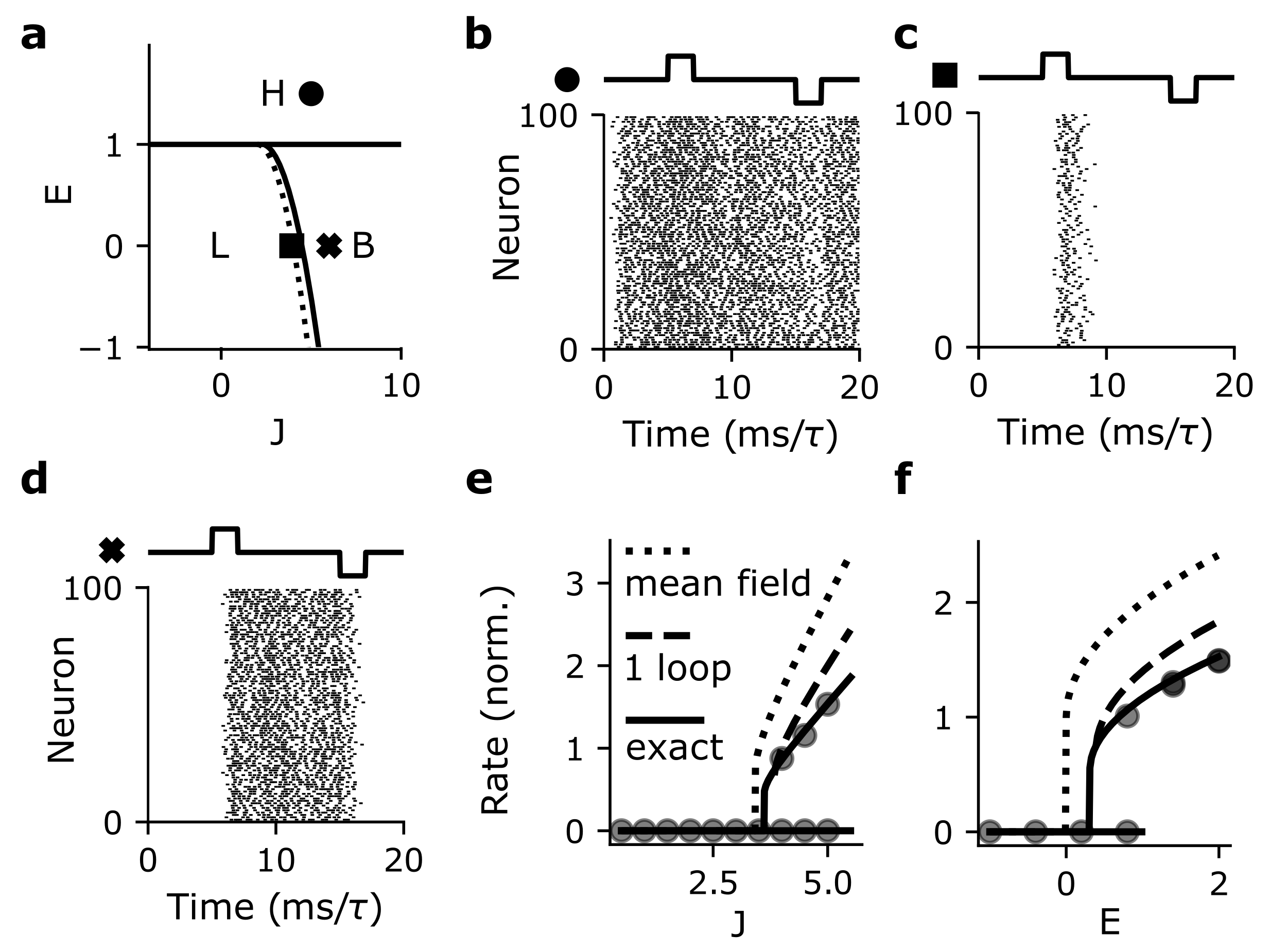}
\caption{Bistable activity in homogenous networks. {\bf a)} Phase diagram of the mean field theory, Eq.~\ref{eq:lif_mft_1pop}, in the input ($E$) vs coupling ($J$) plane. There are three possible states: low activity (L), high activity (H), and bistability (B).  {\bf b-d)} Raster plots of a homogenous networks activity at the parameter locations marked in panel a. At $t=5$ and $t=15$, perturbations of amplitude 2 and duration 2 are applied to the drive $E$ (top). {\bf e)} Bifurcation curve in $J$ with $E=1/2$. {\bf f)} Bifurcation plot in $E$ with $J=4$. Grey circles: simulation. Black dashed: the mean field theory of Eq.~\ref{eq:lif_mft_1pop}. Black solid: the exact rate of the disorder-averaged system, using the numerical self-consistent solution of Eq.~\ref{eq:lif_1pop_exact}. The simulated network has Erd\H{o}-R\`enyi connectivity. Simulated network parameters: $N=100$, $p=0.5$. All non-zero connections have the same weight, $J / (p N)$.
}
\label{fig:bistable} 
\end{figure*}

With a threshold-linear $f$, $f(v) = \lfloor v-1 \rfloor_+$, and pulse coupling, $J(s) = J \delta(s)$, there are three possible steady states of Eq.~\ref{eq:lif_mft_1pop}. The first is $\bar{v}=E$, which exists if $E < 1$. There are two other possible steady states at $v>1$,
\begin{equation} \begin{aligned} \label{eq:lif_mft_fp}
\bar{v}_\pm =& \frac{J \pm \sqrt{J^2 + 4 (E-J)}}{2} \\
\bar{n}_\pm =& \lfloor \bar{v}_\pm - 1 \rfloor_+ ,
\end{aligned} \end{equation}
which both exist if 
\begin{equation} \begin{aligned} \label{eq:lif_mft_fp_bif}
E &< 1 \; \mathrm{and} \;J > 2 + 2 \sqrt{1-E}.
\end{aligned} \end{equation}
Whenever it exists, $\bar{v}_-$ ($\bar{v}_+$) is unstable (stable). If $E>1$, only $\bar{v}_+$ exists. With $J > 2$ and $\frac{J\left(4-J\right)}{4} \leq E < 1$, both steady states exist and the firing rates are thus bistable, with $\bar{v}_-$ providing a separatrix between the attractors $v \rightarrow E$ and $v \rightarrow \bar{v}_+$. The mean field theory has two saddle node bifurcation curves, where the unstable fixed point $\bar{v}_-$ meets either $\bar{v}=E$ or $\bar{v}_+$ (Fig. \ref{fig:bistable}a). 

These bifurcations also appear in the underlying stochastic spiking model. We simulated a network of 100 stochastic LIF neurons (Eq.~\ref{eq:lif}) with Erd\H{o}s-R\'enyi connectivity ($p=0.5$) with different values of the baseline drive $E$ and coupling strength $J$ (marked in Fig. \ref{fig:bistable}a). At times 5 and 15, we applied pulse perturbations to the baseline drive and observed monostable or bistable behavior matching the predictions of the phase diagram (Fig. \ref{fig:bistable}b--d).

The mean field theory neglects all fluctuations in the spiking activity. Due to the nonlinear spike-voltage coupling imparted by the reset mechanism, those fluctuations can impact the firing rate. To determine the magnitude of fluctuation corrections, we computed bifurcation diagrams of the exact firing rate (Eq.~\ref{eq:lif_1pop_exact}; Fig. \ref{fig:bistable}e, f).
The mean field theory systematically overestimates the true firing rates. This implies that fluctuations in the activity suppress firing. 

Similarly to the uncoupled neuron, the impact of fluctuations can be explicitly described by loop corrections to the mean field dynamics (Eq.~\ref{eq:feynman_1loop_1pop}). To one loop, equilibria of $\bar{v}$ are $\bar{v} = E$ (if $\bar{v} < 1$) and 
\begin{equation} \begin{aligned} \label{eq:lif_1loop_fp}
\bar{v}_\pm =& \frac{1}{10} \left(1+4J \pm \sqrt{1+80E+8J(2J-9)} \right)\\
\bar{n}_\pm =& \lfloor \bar{v}_\pm - 1 \rfloor_+
\end{aligned} \end{equation} 
if $\bar{v} > 1$. At one loop, both equilibria of $\bar{v}$ exist if 
\begin{equation} \begin{aligned} \label{eq:lif_1loop_fp_bif}
E &< 1 \; \mathrm{and} \;J > \frac{9}{4} + \sqrt{5(1-E)}.
\end{aligned} \end{equation}

In the model with linear resets and a threshold-linear intensity function, the mean field theory is linear in both the sub- and supra-threshold regimes and does not exhibit bistability.
The classic activity equations can have bistable regimes so long as the nonlinearity saturates, e.g.,~\cite{wilson_excitatory_1972}.
Here, bistability is due to the nonlinear coupling between the spiking and membrane voltage.

The stochastic spiking network may not exhibit true bistability in the bistable regime of the deterministic mean field or one-loop approximations.
Rather, the quiescent state should be truly stable, while the active state is metastable. Fluctuations in the spiking activity may drive the network into the quiescent state. In the quiescent state, there are no fluctuations since all $n$-point correlation functions are sourced by the intensity $f(v)$, which we took to be 0 for $v < 1$. 
If the nonlinearity $f(v)$ were small but finite for $v < 0$, then fluctuations could be maintained in the quiescent state and both would be metastable. The slope of the intensity function at threshold can also play a key role in metastability of the population activity~\cite{robert_dynamics_2016}.

\section{Multiple cell types}
Biological neural networks are composed of diverse types of neurons with cell-type-specific connectivity, e.g.,~\cite{rudy_three_2011, tasic_adult_2016, yao_taxonomy_2021, callaway_multimodal_2021, pfeffer_inhibition_2013, tremblay_gabaergic_2016, seeman_sparse_2018, hage_synaptic_2022}. Motivated by this, we consider a network with $M$ populations, which impose a block structure on the connectivity matrix $\bm{J}$. 
The average over the connectivity proceeds as for the single population, with an order parameter for each population's mean activity. This yields a $M$-dimensional mean field theory. In the large-$N$ limit, the partition functional is
\begin{widetext}
\begin{equation} \begin{aligned} \label{eq:saddle_point_Z_ei_homog}
Z^* = \int \mathcal{D} \bm{v}  \int \mathcal{D} \dot{\bm{n}}  \int \mathcal{D} \tilde{\bm{v}}  \int \mathcal{D} \tilde{\bm{n}} \; \exp \sum_{\alpha=1}^M \Bigg(& -\tilde{v}_\alpha^T \left( \partial_t v_\alpha + v_\alpha - E_\alpha - \sum_{\beta=1}^M J_{\alpha \beta} \ast \langle \dot{n}_\beta \rangle + \dot{n}_\alpha v_\alpha \right) \\
&- \tilde{n}_\alpha^T \dot{n}_\alpha + \left(e^{\tilde{n}_\alpha}-1 \right)^T f(v_\alpha) \Bigg).
\end{aligned} \end{equation}
\end{widetext}
For self-averaging networks, the density factorizes over the populations and neurons so the neurons again spike independently given a self-consistent mean field input.
 The typical spike train of population $\alpha$ ($\alpha \in [M]$) is an inhomogenous Poisson process. 
 If the population-averaged activities $\langle \dot{n}_\alpha \rangle$ are constant in time, the mean first passage times are
\begin{equation} \begin{aligned} \label{eq:lif_multi-pop_exact}
\langle s_\alpha \rangle 
=& \ln \left( \frac{C_\alpha}{C_\alpha-1} \right) + \left(\frac{C_\alpha-1}{e} \right)^{1-C_\alpha} \gamma(C_\alpha-1, C_\alpha-1),
\end{aligned} \end{equation} 
where $C_\alpha = E + \sum_{\beta=1}^M J_{\alpha \beta}  \langle \dot{n}_\beta \rangle$. In a stationary state, the rate is $\langle \dot{n}_\alpha \rangle = 1 / \langle s_\alpha \rangle$.
The mean field approximation of the membrane voltages is
\begin{equation} \label{eq:lif_mft_multi_pop}
\partial_t v_\alpha = - v_\alpha - v_\alpha f(v_\alpha) + E_\alpha + \sum_\beta J_{\alpha \beta} \ast f(v_\beta) .
\end{equation}
The one-loop equations of motion, similarly, are given by accounting for the input across populations in  Eq.~\ref{eq:feynman_1loop_1pop}.

\section{Bistable activity in excitatory-inhibitory networks}
\begin{figure}[ht!] \includegraphics[]{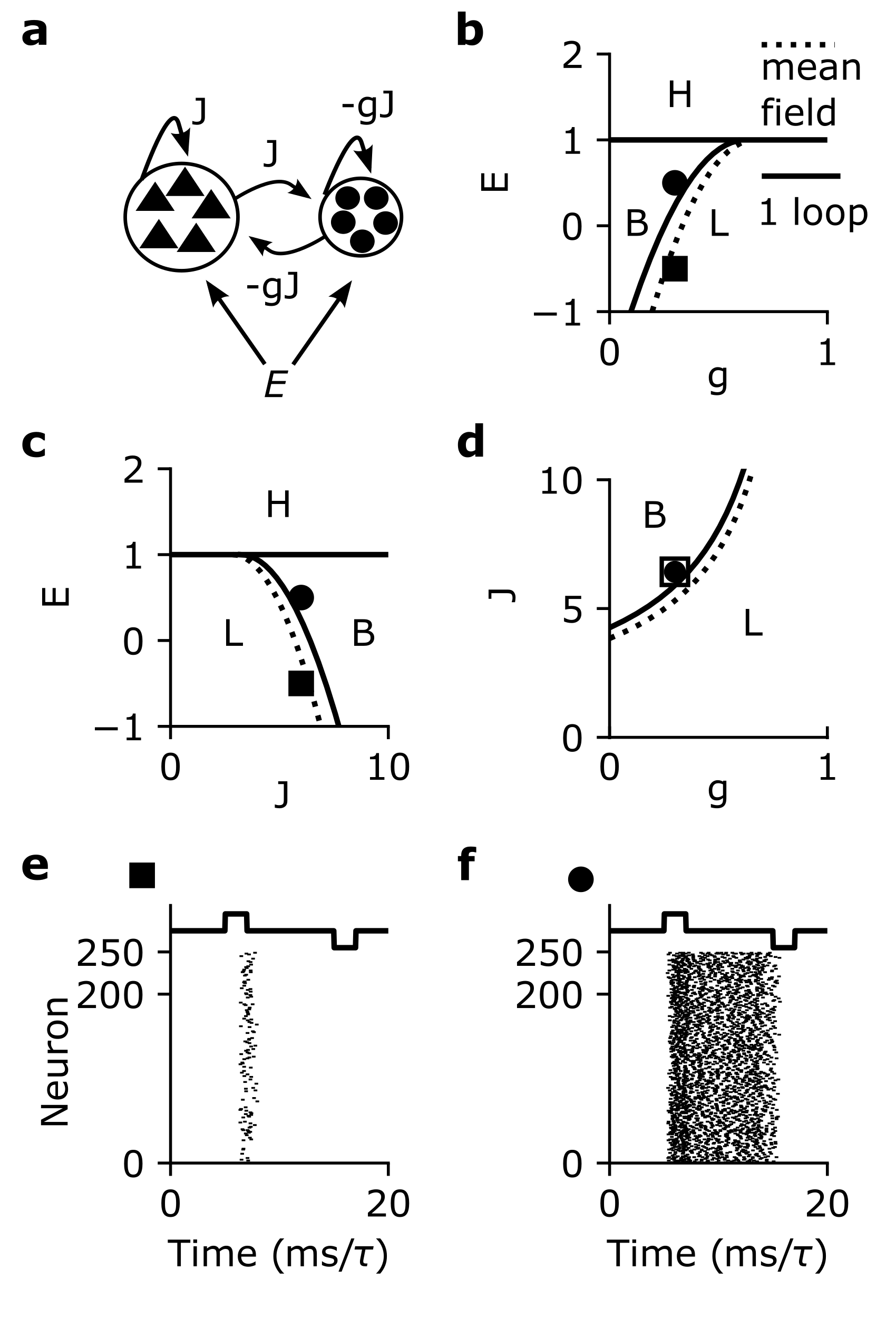}
\caption{Bistable activity in excitatory-inhibitory networks. {\bf a)} Network diagram. {\bf b)} Phase diagram in the input ($E$) vs inhibitory strength ($g$) plane with $J=6$. There are three possible states: low activity (L), high activity (H), and bistability (B). {\bf c)} Phase diagram of the two-dimensional mean field theory, Eq.~\ref{eq:lif_mft_multi_pop}, in the input ($E$) vs coupling strength ($J$) plane with $g=0.3$. {\bf d)} Phase diagram in the coupling vs inhibitory strength plane with $E=0.5$. {\bf e, f)} Example simulations with $(J,g)=(6, 0.3)$, with $E=-0.5$ ({\bf e}) or $E=0.5$ ({\bf f}). The network has a block-Er\H{o}s-R\`enyi structure. Simulated network parameters: population sizes $(N_e, N_i)=(200, 50)$, excitatory output connection probabilities $p_{ee} = p_{ie}=0.2$, inhibitory output connection probabilities $p_{ei} = p_{ii} = 0.8$. Within each block, all non-zero connections have the same weight, e.g., $J/(N_e p_e)$ for non-zero excitatory projections.
}
\label{fig:bistable_ei} 
\end{figure}

Here, we consider the classic excitatory-inhibitory network with pulse coupling and mean connection strengths
\begin{equation} \label{eq:ei_mean_J}
\begin{pmatrix}
J_{EE} & J_{EI} \\
J_{IE} & J_{II}
\end{pmatrix}
=
\begin{pmatrix}
J & -gJ \\
J & -gJ
\end{pmatrix},
\end{equation}
as in~\cite{amit_model_1997, brunel_dynamics_2000} (Fig. \ref{fig:bistable_ei}a). 
With input $E$ to both populations, the mean rates of the excitatory and inhibitory populations are equal since they receive the same external and recurrent inputs. 
The self-consistent fixed points with positive rates are the same as those in the single-population network with the replacement $J \rightarrow J(1-g)$ (Eq.~\ref{eq:lif_mft_fp} for the mean field theory, Eq.~\ref{eq:lif_1loop_fp} to one loop). 
In the mean field theory, both fixed points exist if
\begin{equation} \begin{aligned}
E &< 1 \; \mathrm{and} \; J > 2 \; \mathrm{and} \\
g & \leq
 1 - \frac{2}{J}\left(1 + \sqrt{1-E}\right).
\end{aligned} \end{equation}
Here we have highlighted the requirement for $g$ as a function of the other model parameters; $J > 2$ is a necessary condition for bistability in the mean field theory, but depending on the values of $g$ and $E$ the greatest lower bound for $J$ may be above $2$.
With both population voltages under threshold, there is the stable fixed point $\bar{v} = E$, if $E \leq 1$. 
If $E>1$, only $\bar{v}_+$ exists. 
The Jacobian eigenvalue $J(1-g)-2v$ is positive for $\bar{v}_-$ and negative for $\bar{v}_+$ root. 
So if these fixed points exist, the one at higher $v$ is stable and the other a saddle. 
Similarly, in the one-loop theory both fixed points exist if
\begin{equation} \begin{aligned}
E &< 1 \; \mathrm{and} \; J >  \frac{9}{4}  \; \mathrm{and} \\
g & \leq \; 1 - \frac{1}{J} \left(\frac{9}{4} - \sqrt{5(1-E)} \right) 
\end{aligned} \end{equation}
As for the single-population network, the existence conditions for these fixed points define saddle node bifurcation curves for the mean field and one-loop theories (Fig. \ref{fig:bistable_ei}b-d). If the inhibitory coupling strength is sufficiently low, we have the same types of bifurcation curves as in the single-population network (Fig. \ref{fig:bistable_ei}c). If the inhibitory coupling $g$ is too strong, the only stable equilibrium is the low-rate state (Fig. \ref{fig:bistable_ei}b, d). 

These bifurcations also appear in the stochastic spiking network with block-Erd\H{o}s-R\'enyi connectivity (Fig. \ref{fig:bistable_ei}e, f; network parameters are given in the caption). 


\section{Fluctuations} \label{sec:fluctuations}
\begin{figure}[] \includegraphics[]{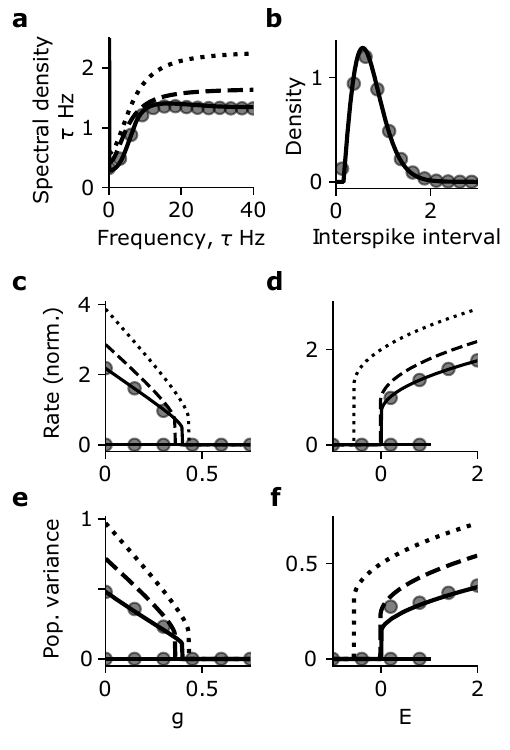}
\caption{Fluctuations in an excitatory-inhibitory network with symmetric external inputs, $E_e=E_i=E$. {\bf a)} Spike train power spectrum with $(J, g, E) = (6, 0.3, 1.2)$. Dots: simulation of a network with 200 excitatory and 50 inhibitory neurons (population-averaged power spectrum). Dotted: the perturbative tree-level approximation (expanded around the deterministic mean field theory). Dashed: the tree-level approximation around the one-loop rates. Solid: the renewal prediction of Eq.~\ref{eq:spectrum_exact}. 
{\bf b)} Interspike interval density with $(J, g, E) = (6, 0.3, 1.2)$. Dots: simulation. Solid: the renewal prediction of Eq.~\ref{eq:fpt}.   
{\bf c, d)} Bifurcation diagrams for firing rate as a function. 
{\bf e, f} Bifurcation diagrams for spike train variance.
In ({\bf c}, {\bf e}), $(J,E)=(6, 0.5)$. In ({\bf d}, {\bf f}), $(J, g)=(6, 0.25)$. 
Simulated network parameters: population sizes $(N_e, N_i)=(200, 50)$, excitatory output connection probabilities $p_{ee} = p_{ie}=0.5$, inhibitory output connection probabilities $p_{ei} = p_{ii} = 0.8$.
}
\label{fig:corrs_ei} 
\end{figure}

The temporal structure of fluctuations can shape sensory codes~\cite{chacron_population_2008, yu_membrane_2010, giridhar_timescale-dependent_2011} and determine neural circuit structures through spike timing-dependent plasticity~\cite{kempter_hebbian_1999, gilson_stdp_2010, babadi_pairwise_2013, ocker_self-organization_2015, tannenbaum_shaping_2016, ocker_training_2018, montangie_autonomous_2020}. 
The classic Fokker-Planck mean field theory of integrate-and-fire networks assumes that the membrane voltages experience a white Gaussian noise ~\cite{amit_model_1997, brunel_dynamics_2000, lindner_transmission_2001}. The resulting predictions for the spike trains' power spectra are not white, however, so these predictions are not self-consistent~\cite{lindner_superposition_2006}. In the stochastic integrate-and-fire model, the output spike trains also are not white. In the excitatory-inhibitory network, for example, the population-averaged power spectrum exhibits a high-pass shape with a slight resonance (Fig.~\ref{fig:corrs_ei}a, dots). This is similar to the shape of the power spectrum of networks of deterministic integrate-and-fire neurons with white noise inputs~\cite{doiron_oscillatory_2004}. The one-loop equations of motion account for Gaussian fluctuations, but do not make any assumptions about their temporal structure. 
We will next discuss the temporal structure of fluctuations in this Gaussian approximation and the full prediction from renewal theory.

The exact mean field theory, Eq.~\ref{eq:saddle_point_Z_ei_homog}, is of an inhomogeneous Poisson process receiving the self-consistent mean input $\sum_\beta J_{\alpha \beta} \langle \dot{n}_\beta \rangle$. 
Substituting this into Eq.~\ref{eq:lif_postspike}, yields the post-spike membrane voltage, which defines the intensities $f(v_\alpha(t))$. 
The interspike interval density is
$p(s_\alpha) = f(v_\alpha(s)) \; \exp \left(-\int_0^s dt \; f(v_\alpha(t)) \right).$
For a threshold-linear intensity function, it is
\begin{equation} \label{eq:fpt}
p(s_\alpha) = \begin{cases} &0, \; s_\alpha \leq \ln \frac{C_\alpha}{C_\alpha - 1} \\ \\
& \left( C_\alpha \left( 1 - e^{-s_\alpha} \right) - 1 \right) \\
& \times \exp - \left( C_\alpha e^{-s_\alpha} + \left(C_\alpha -1 \right) \left(s_\alpha - 1 - \ln \frac{C_\alpha}{C_\alpha - 1} \right) \right), \\
& s_\alpha > \ln \frac{C_\alpha}{C_\alpha - 1}
\end{cases}
\end{equation}
where $C_\alpha = E + \sum_\beta J_{\alpha \beta} \langle \dot{n}_\beta \rangle$ and $\gamma$ is again the lower incomplete gamma function.
This provides an exact prediction for the interspike interval density in the limit $N \rightarrow \infty$, accurate for populations of a few hundred neurons (Fig. \ref{fig:corrs_ei}b).
The interspike interval distribution defines the spike train power spectrum $C(\omega)$ of a renewal process~\cite{cox_renewal_1962}:
\begin{equation} \label{eq:spectrum_exact}
C(\omega) = \langle \dot{n} \rangle \frac{1 - \lvert p(\omega) \rvert^2 }{\lvert 1 - p(\omega) \rvert^2 }.
\end{equation}
Together, Eqs.~\ref{eq:fpt} and \ref{eq:spectrum_exact} provide an exact prediction for the typical power spectrum in a large homogenous network. Computing the Fourier transform $p(\omega)$ numerically, we see that these predictions are quantitatively accurate in simulations of a few hundred neurons (Fig. \ref{fig:corrs_ei}a, dots vs solid). 

For analytic approximations of the power spectrum, we turn to the field theoretic formulation.
If the fluctuations or the nonlinearity are weak, we can expand the density pertubatively around a solution of the deterministic mean field theory (Appendix~\ref{app:perturb}).
The connected two-point function of the spike trains can then be calculated diagrammatically. With a threshold-linear intensity function,
\begin{equation} \begin{aligned} \label{eq:2pt_expansion}
\llrrangle{\dot{n}^2} =&
\vcenter{ 
\hbox{\begin{tikzpicture} \begin{feynman}[small]
\vertex (a) at (0, 1) {};
\vertex (b) at (0, 0) {};
\vertex [dot] (c) at (.5, .5) {};
\diagram*{
(a) -- (c) -- (b);
}; 
\end{feynman} \end{tikzpicture} } } 
+
\vcenter{
\hbox{\begin{tikzpicture} \begin{feynman}[small]
\vertex (a) at (0, 1) {};
\vertex (b) at (0, 0) {};
\vertex [empty dot] (c) at (.5, .5) {};
\vertex [empty dot] (d) at (1, .5) {};
\vertex [dot] (e) at (1.5, .5) {};
\diagram*{
(a) -- (c) -- (b);
(c) --[dashed] (d);
(d) --[half right] (e) --[boson, half right] (d);
};
\end{feynman} \end{tikzpicture} } }
+
\vcenter{ 
\hbox{\begin{tikzpicture} \begin{feynman}[small]
\vertex (a) at (0, 1) {};
\vertex [empty dot] (b) at (.5, .75) {};
\vertex [empty dot] (c) at (1, .5) {};
\vertex (d) at (0, -.25) {};
\vertex [dot] (e) at (1.5, .25) {};
\diagram*{
(a) --[gluon] (b);
(b) --[half right] (c);
(b) --[boson, half left] (c);
(c) --[boson] (e);
(d) -- (e);
}; \end{feynman} \end{tikzpicture} } } 
+ \mathcal{O}(2 \; \mathrm{loops}).
\end{aligned} \end{equation}
The expansion may contain terms with up to infinitely many loops, inducing dependence on $n$-point correlation functions of all orders. With an intensity function non-linear at the mean voltage there would be additional diagrams, containing internal vertices with multiple incoming $\vcenter{ \hbox{
\feynmandiagram[small, horizontal=a to b, inline=(a.base)]{
a --[photon] b [],
 }; }}$ edges.
The same is true for any cumulant of the activity. 
The simplest approximation of the two-point correlation is the tree-level approximation given by the first diagram of Eq.~\ref{eq:2pt_expansion}, 
\begin{equation} \begin{aligned} \label{eq:2pt_tree}
\llrrangle{\dot{n}^2_\alpha}_0(\omega) =& \frac{\bar{v}_\alpha^2 + \omega^2}{4 \bar{v}_\alpha^2 + \omega^2} f(\bar{v}_\alpha),
\end{aligned} \end{equation}
where $\bar{v}_\alpha$ is a solution to the mean field equation for population $\alpha$. 
At $\omega = 0$, this yields $\llrrangle{\dot{n}^2_\alpha}(0) \approx f(\bar{v}_\alpha) / 4$. For $\omega \rightarrow \infty$, $\llrrangle{\dot{n}^2}(\omega) \rightarrow f(\bar{v}_\alpha)$, the mean-field approximation to the intensity. 
This simple approximation captures the high-pass nature of the power spectrum but is not quantitatively accurate (Fig.~\ref{fig:corrs_ei}a, dotted line).

The one-loop predictions for the mean membrane voltage and rate account for second-order fluctuations to tree level. 
For the spike train power spectrum this again corresponds to Eq.~\ref{eq:2pt_tree}, but with $\bar{v}$ a solution to the one-loop equations of motion.
This provides a more accurate prediction of the power spectrum (Fig.~\ref{fig:corrs_ei}a, dashed line) due to the improved estimate of the intensity, $f(\bar{v}_\alpha)$.

As the coupling or input strength brings the network to a bifurcation, the spike train variance $\llrrangle{\dot{n}^2_\alpha}_0(0)$ undergoes a sharp transition from 0 in the quiescent state to positive values in the active state (Fig.~\ref{fig:corrs_ei}c-f). The transition in the spike train variance follows that in the rate, since all correlation functions are sourced by the intensity $f(\bar{v})$.

\section{Inhibitory stabilization} \label{sec:isn}
In recent years, a body of work has emerged suggesting that mammalian cortices resides in an inhibition-stabilized regime~\cite{griffith_stability_1963, wilson_excitatory_1972, ozeki_inhibitory_2009, kato_network-level_2017, adesnik_synaptic_2017, sanzeni_inhibition_2020}. 
There are two requirements for an excitatory-inhibitory network to be inhibition-stabilized: the network must occupy a stable attractor, but the excitatory population would be unstable on its own. 
These are difficult to directly test experimentally. 
Fortunately, inhibition-stabilized fixed points have another signature: paradoxical responses to inhibitory neuron stimulation. 
In an inhibition-stabilized network, stimulation of the inhibitory neurons leads to a paradoxical reduction of their firing rates~\cite{tsodyks_paradoxical_1997}. If there are multiple inhibitory subtypes, the net inhibitory input to excitatory neurons decreases upon inhibitory neuron stimulation~\cite{litwin-kumar_inhibitory_2016}. The widespread experimental observation of paradoxical responses, and other response patterns consistent with inhibition-stabilized networks, raises the question: is inhibitory stabilization a generic property, or does it require fine-tuned parameters towards which cortical networks develop?

The inhibition-stabilized regime, and paradoxical responses as its signature, are predictions of the classic activity equations, Eq.~\ref{eq:amari}. 
Does an inhibition-stabilized regime exist in the mean field theory of Eqs.~\ref{eq:lif_mft_multi_pop} and~\ref{eq:ei_mean_J}? 
The stability requirements are determined from the Jacobian matrix, 
\begin{equation}
\begin{pmatrix}
-1 - \bar{f}_e + (J- \bar{v}_e) f^{(1)}_e & -g J f^{(1)}_i \\
J f^{(1)}_e & -1 - f_i - (g J + \bar{v}_i)f^{(1)}_i
\end{pmatrix} ,
\end{equation}
where $\bar{f}_\alpha = f(\bar{v}_\alpha)$ and $f^{(1)}_\alpha = \frac{d}{dv} f(v) \vert_{\bar{v}_\alpha}$.
For a fixed point to be inhibition-stabilized, the first element of its Jacobian must be positive (the excitatory-only subnetwork would be unstable), but the maximum real part of its eigenvalues negative (the full network is stable). 
%
For the threshold-linear intensity function, $f^{(1)}(\bar{v}_\alpha) = \theta(\bar{v}_\alpha - 1)$, where $\theta(x)$ is the Heaviside step function. This leads to the requirement that for the excitatory subnetwork to be linearly unstable with a positive firing rate, $1 < \bar{v}_E < J/2$. 

\begin{figure} \includegraphics[]{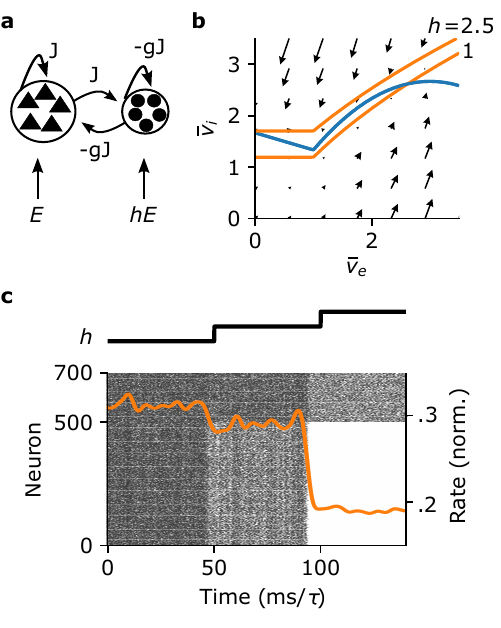}
\caption{Paradoxical responses to inhibitory stimulation.  {\bf a)} Excitatory-inhibitory network with asymmetric drive. {\bf b)} Phase diagram and nullclines of the excitatory (blue) and inhibitory (orange) firing rates for the excitatory-inhibitory network with threshold-linear rate functions. {\bf c)} Simulation of a block-Erd\H{o}s-R\'enyi network with $p_{ee} = p_{ie} = 0.5$, $p_{ei} = p_{ii} = 0.8$. At time 0, $(E, h) = (2, 1)$. At times 50 and 100, $h$ increases by $\frac{3}{4}$. Orange: inhibitory population-averaged spike train, smoothed with a Gaussian kernel of width 2 for visualization. Parameters for {\bf b}, {\bf c}: $(J, g, E) = (6, 1/2, 2)$. Simulated block Erd\H{o}s-R\`enyi network parameters as in Fig.~\ref{fig:bistable_ei}.
}
\label{fig:paradox} 
\end{figure}

Do paradoxical responses to inhibitory stimulation occur in the stochastic LIF network? To investigate this, we return to the tractable threshold-linear intensity function.
We allow the external input to vary between the two populations, $\bm{E} = (E, hE)$ (Fig. \ref{fig:paradox}a). $h$ controls the relative strength of the input to the inhibitory population.  When both population voltages are above threshold, the mean field inhibitory and excitatory nullclines are at
\begin{equation} \begin{aligned} \label{eq:ei_nullclines}
\bar{v}_i^* &= \frac{-gJ + \sqrt{g^2 J^2 + 4(J(\bar{v}_e - 1 + g) + hE)}}{2} \\
\bar{v}_i &= \left(-(\bar{v}_e^*)^2 + J \bar{v}_e^* - J(1-g) + E \right) / gJ .
\end{aligned} \end{equation}
$\bar{v}_\alpha^*$ is the nullcline of population $\alpha \in \{e, i\}$. The supra-threshold inflection point of the excitatory nullcline is at $\bar{v}_e = J/2$. An inhibition-stabilized fixed point must thus be on the increasing branch of the excitatory nullcline.
$h$ does not affect the excitatory nullcline but shifts the inhibitory nullcline. 
An increase in $h$ will lead to a paradoxical reduction in firing rates if it shifts a stable fixed point to lower $\bar{v}_i$. 
For example, consider the case when there is a single fixed point on the increasing side of the excitatory nullcline, to the left of its peak (Fig. \ref{fig:paradox}b). 
Increasing $h$ shifts the inhibitory nullcline up and to the left, moving that fixed point to a lower $(\bar{v}_e, \bar{v}_i)$. Depending on the magnitude of the shift, it may also take the dynamics through a bifurcation into a bistable regime. A sufficiently large increase in $h$ can shift the network into a regime with no excitatory activity, which can also lead to a net decrease in inhibitory rates (Fig. \ref{fig:paradox}b, c). 

In what regions of parameter space does an inhibition-stabilized fixed point exist? As discussed above, for the excitatory subnetwork to be unstable, with non-zero excitatory rate, requires that the fixed point be on the middle branch of the excitatory nullcline: $1 < \bar{v}_E < J/2$. 
The inhibitory nullcline is an increasing function of $\bar{v}_e$. The excitatory nullcline increases for $\bar{v}_e^*$ close to 1 and decreases for sufficiently large $\bar{v}_e^*$. 
At threshold ($\bar{v}_e=1$) the inhibitory nullcline must be below the excitatory nullcline:
\begin{equation} \label{eq:ei_single_fp}
\sqrt{(gJ)^2 + 4gJ + 4 h E} < \frac{2}{gJ} \left(E + gJ - 1\right) + gJ .
\end{equation}
If $h=1$, this requirement imposes that $E > 1$; at $E=1$ the two sides are equal, and the difference of the two sides grows as $\sqrt{E}$. 

\begin{figure} \includegraphics[]{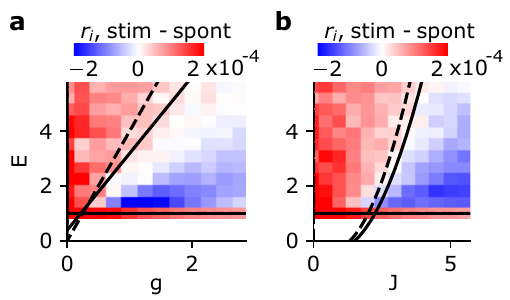}
\caption{Phase diagrams for paradoxical responses to inhibitory stimulation.  {\bf a)} Boundaries of the paradoxical response region with $J=4$. Dashed line: mean field theory, Eq.~\ref{eq:ei_paradoxical}. Solid line: one-loop theory, Eq.~\ref{eq:ei_paradoxical_1loop}. Color: simulation. Each simulation lasts for 200 time units; at time 100, the inhibitory drive switches from $hE = E$ to $hE = E + 0.1$. {\bf b)} As in {\bf a}, with $g=2$. Simulated block Erd\H{o}s-R\`enyi network parameters as in Fig.~\ref{fig:bistable_ei}.
}
\label{fig:paradox_region} 
\end{figure}

The peak of the excitatory nullcline is at $\bar{v}^*_e = J/2$. At the peak of the excitatory nullcline, $\bar{v}_i = \left(J^2 / 4 - J(1-g) + E \right) / gJ$. At $\bar{v}_e = J/2$, the inhibitory nullcline should be above the excitatory nullcline: 
\begin{equation} \begin{aligned} \label{eq:ei_paradoxical}
&\sqrt{(gJ)^2 + 4gJ + 2 J (J-2) + 4 h E } \\
&> \frac{2}{gJ} \left( \frac{J^2}{4} - J(1-g) + E \right) + gJ.
\end{aligned} \end{equation}
Together, Eqs.~\ref{eq:ei_single_fp} and \ref{eq:ei_paradoxical} provide sufficient conditions for a paradoxical response to inhibitory stimulation in the mean field theory. At fixed drive $E$, they predict a paradoxical response for sufficiently large $J$ or $g$. For stronger $E$, these minimal couplings increase (Fig. \ref{fig:paradox_region}a, b, dashed).

To estimate how fluctuations impact inhibitory stabilization, we compute the one-loop nullclines (each with $v_i$ as a function of $v_e$):
\begin{equation} \begin{aligned}
\bar{v}_i^* =& \frac{ 1 - 4 g J + \sqrt{1 + 80 h E + 8 g J (9 + 2 g J) + 80 J (v_e - 1)} } { 10} \\
v_i =& \left(4 \left(E - J + gJ\right) + \bar{v}_e^*(4J+1) - 5 (\bar{v}_e^*)^2 \right) / 4 g J
\end{aligned} \end{equation} 
The inflection point of the one-loop $\bar{v}_e$ nullcline is at $\bar{v}_e^* = \left(1 + 4 J\right)/10$.
At one loop, for the inhibitory nullcline to be below the excitatory nullcline at threshold requires
\begin{equation}
10E > 10 + g J (-9 - 4 g J + \sqrt{1 + 80 E h + 8 g J (9 + 2 g J)})
\end{equation} 

for the inhibitory nullcline to be above the excitatory nullcline at $\bar{v}_e = \left(1 + 4 J\right)/10$ requires
\begin{equation} \begin{aligned} \label{eq:ei_paradoxical_1loop}
1 + 80 E <& \, 8 J \Big(9 - 2 J - g \big(9 + 4 g J \\
&-\sqrt{1 + 80 E h + 8 J (-9 + 4 J + g (9 + 2 g J))} \big) \Big)
\end{aligned} \end{equation} 
For fixed $E$ and $J$, this one-loop boundary requires a higher $g$ (stronger inhibition) than the mean field boundary, better matching the transition observed in simulations (Fig.~\ref{fig:paradox_region}a, dashed vs solid).
Similarly, for fixed $E$ and $g$, the one-loop boundary is at higher $J$ (stronger coupling) than the mean field boundary (Fig.~\ref{fig:paradox_region}b, dashed vs solid). 
Together, this comparison indicates that fluctuations shift the region of paradoxical responses to more strongly coupled networks. 
This comports with the role of fluctuations in suppressing activity.

A paradoxical response could also occur from other dynamical regimes than the single fixed point on the decreasing branch of the excitatory nullcline, such as from a bistable regime. 
To test whether the underlying spiking model exhibits paradoxical responses, we simulated excitatory-inhibitory networks while varying $J$ and $g$. 
For each network, we applied a perturbation of amplitude 0.1 to the inhibitory population's input and computed the inhibitory population's average firing rate before and after the perturbation. 
With $J$ fixed and varying $g$, we observed paradoxical responses for sufficiently large $g$ (Fig.~\ref{fig:paradox_region}a, blue). 
Similarly, with $g$ fixed and varying $J$, we observed paradoxical responses for sufficiently large $J$ (Fig. \ref{fig:paradox_region}b, blue). 
The one-loop predictions better match the region of paradoxical responses than the mean field predictions (Fig.~\ref{fig:paradox_region}).

\section{Discussion}
We constructed a path integral representation for the joint probability density functional of the membrane voltage and spike trains of a network of stochastic LIF neurons, Eq.~\ref{eq:lif_action}. 
This exposed a simple deterministic mean field theory for spiking networks: activity equations with an additional rate-dependent leak arising from the spike reseting (Eq.~\ref{eq:lif_mft}). 
It also exposed fluctuation corrections to the mean field theory, arising from the two nonlinearities of the intensity function and spike reset (Eq.~\ref{eq:feynman_1loop_1pop}), the latter of which suppresses activity (Fig.~\ref{fig:intro}).
These $N$-dimensional systems expose predictions for the activity that depend on a particular connectivity ${\bm J}$. 
Large-scale electron microscopy is now revealing such wiring diagrams, e.g.,~\cite{ohyama_multilevel_2015, berck_wiring_2016, wanner_dense_2016, morgan_fuzzy_2016, larderet_organization_2017, hildebrand_whole-brain_2017, eichler_complete_2017, zheng_complete_2018, motta_dense_2018, cook_whole-animal_2019, xu_connectome_2020, consortium_functional_2021, vishwanathan_predicting_2021}. 
Predicting the microscopic dynamics of even deterministic, threshold-linear neuronal networks is challenging~\cite{curto_fixed_2019, curto_combinatorial_2020, santander_nerve_2022, parmelee_sequential_2022}.
Statistical approaches focusing on stochastic models allow the prediction of correlations between specific neurons' activity through linear response or Hawkes theory and its generalizations~\cite{hawkes_spectra_1971, sejnowski_stochastic_1976, ostojic_how_2009, pernice_how_2011, trousdale_impact_2012, jovanovic_interplay_2016, ocker_linking_2017}.
Eq.~\ref{eq:lif_action} provides a starting point for making such predictions in a stochastic LIF network.
 
Here, we instead used the path integral representation to derive a population-averaged stochastic field theory for large networks with homogenous coupling, including multi-population systems like excitatory-inhibitory networks. 
That stochastic field theory was of the form of a renewal process with a self-consistent input (Eqs.~\ref{eq:saddle_point_Z_1pop_homog},~\ref{eq:saddle_point_Z_ei_homog}). 
Robert \& Touboul studied the homogenous stochastic LIF network rigorously ~\cite{robert_dynamics_2016}. 
They proved that the mean field process, Eq.~\ref{eq:saddle_point_Z_1pop_homog}, can have one or several invariant densities depending on the form of the firing function. 
The stochastic field theory admits low-dimensional mean field and loop approximations of the voltage and rate as simple functions of the model's parameters.
Using these, we demonstrated bistability of the deterministic mean field theory and its extension to the stochastic system, and studied the contributions of recurrent inhibition and spike resetting to stabilizing network activity. 
We also found that fluctuations suppress activity through the spike reset also in coupled networks (Figs.~\ref{fig:bistable}e-f,~\ref{fig:corrs_ei}c-e).
Excitatory-inhibitory networks of deterministic integrate-and-fire neurons can also exhibit bistable equilibrium rates if the inhibition is not too strong~\cite{brunel_dynamics_2000, caceres_analysis_2011}.
The field-theoretic description here does not rely on a white noise approximation for the membrane voltages, but exposes a systematic method for calculating their statistics.
It requires here, however, a model with stochastic spike emission. 

Deterministic integrate-and-fire networks can also exhibit spatial, temporal and spatiotemporal transitions~\cite{brunel_dynamics_2000, laing_stationary_2001, rosenbaum_balanced_2014}. Temporal, spatial and spatiotemporal bifurcations are often understood through the classic activity equations~\cite{bressloff_spatiotemporal_2011}. The field theory developed here provides a route to uncovering bifurcations in networks of stochastic integrate-and-fire neurons with more temporal or spatial structure in their interactions, as well as investigating the impact of spiking fluctuations on such transitions.

In the classic activity equations (e.g., Eq.~\ref{eq:amari}), recurrent inhibition is necessary to stabilize strongly coupled networks~\cite{griffith_stability_1963, wilson_excitatory_1972}.
An inhibition-stabilized regime can be exposed by a paradoxical reduction of inhibitory activity after inhibitory stimulation~\cite{tsodyks_paradoxical_1997}. 
We calculated the phase diagram for paradoxical responses in stochastic LIF networks, and found that an inhibition-stabilized regime exists in wide regions of parameter space (Figs.~\ref{fig:paradox},~\ref{fig:paradox_region}). This suggests a generic mechanism underlying the observation of paradoxical responses widely in mammalian cortex~\cite{ozeki_inhibitory_2009, kato_network-level_2017, adesnik_synaptic_2017, sanzeni_inhibition_2020}.

There are two complementary approaches to our focus on the density functional of sample paths, $p[\bm{v}(t), \dot{\bm n}(t)]$, for the stochastic LIF model. 
These complementary approaches focus on the time-dependent probability density function of the membrane voltages, $p(v, t)$, across a population of neurons~\cite{knight_dynamics_1972}. In the $N \rightarrow \infty$ limit and with $J_{ij} \sim 1/N$, the population density of membrane voltages in a stochastic LIF network obeys a Volterra integral equation~\cite{gerstner_time_1995, gerstner_population_2000}. 
That integral equation can also be written as a partial differential equation, which rigorously exposes the stochastic stability of the population densities in a mean field limit~\cite{de_masi_hydrodynamic_2015, fournier_toy_2016, duarte_model_2016, cormier_long_2020}.
A finite-size analysis introduces a stochastic term to the population density equations~\cite{schwalger_towards_2017, schmutz_finite-size_2021}. Alternatively, moments for finite-size networks can be analyzed through a replica mean-field approach~\cite{baccelli_replica-mean-field_2019, baccelli_pair-replica-mean-field_2021, yu_metastable_2022}. The path integral approach also exposes a finite-size mean field theory (Eq.~\ref{eq:lif_mft}). Fluctuation corrections to that finite-$N$ mean field theory can be obtained in the same way as for the large-$N$, connectivity-averaged system.

The field theoretic approach is practical and flexible. It exposes simple analytic approximations for any cumulant of the membrane voltages and/or spike trains via diagrammatic methods, is amenable to finite-size corrections, and applies readily to other models such as those with temporal synaptic interactions, spatially dependent connectivity, conductance-based or strong $\mathcal{O}(1/\sqrt{N})$ synapses, and additional nonlinearities in the single-neuron dynamics. 

A previous version of this article, now retracted \url{https://journals.aps.org/prx/abstract/10.1103/PhysRevX.12.041007}, contained errors described in the retraction notice \url{https://journals.aps.org/prx/abstract/10.1103/PhysRevX.13.029904}. These errors have been corrected in the current article. Related code can be found at~\cite{ocker_github}.
\bigskip

\section{Acknowledgements}
I thank Michael A. Buice, Brent Doiron, and Stefano Recanatesi for helpful feedback.
\medskip
\appendix
\section{Joint probability density functional} \label{app:density}
We will construct the joint probability density of the membrane voltages and spike trains using the response variable path integral formalism~\cite{martin_statistical_1973, dominicis_techniques_1976, janssen_lagrangean_1976, jensen_functional_1981}, reviewed in~\cite{chow_path_2015, hertz_path_2016, helias_statistical_2020}. We will use boldface lowercase variables for vectors and boldface capital letters for matrices/operators. Given the membrane voltages $v_{it}$, we will require that the spikes generated in the network are conditionally independent across neurons $i$ and time points $t$. Here we take the model to be already non-dimensionalized, so that time is measured in units of the membrane time constant $\tau$ and the voltage resets to 0 after a spike.
The joint probability density of the membrane voltages $\bm{v}$ and the spikes $d\bm{n}$, conditioned on the stochastic spike generation, is
\begin{widetext}
\begin{equation} \begin{aligned} 
p(\bm{v}, \bm{n} \vert {\bf \eta}) = \prod_{i=1}^N \prod_{t=1}^{T-1} & \delta \left(\frac{dv_{it}}{dt} + v_{it} + \frac{dn_{i,t-1}}{dt}v_{it}  - E_{it} - \sum_{j, s} J_{ijs} \, dn_{j,t-s} \right) \times \delta \left(dn_{it} - \eta_{it} \right).
\end{aligned} \end{equation}
\end{widetext}
Here, $\eta_{it} \sim \mathrm{Bernoulli}\left(f(v_{it}) \, dt\right)$. Introducing the Fourier representation of the delta functions and marginalizing over ${\bf \eta}$ yields the joint density
\begin{widetext}
\begin{equation} \begin{aligned}
p(\bm{v}, \bm{n}) = \int D\tilde{\bm{v}} \int D \tilde{\bm{n}} \; \exp \Bigg(& \sum_{i, t} -\tilde{v}_{it} \left(\frac{dv_{it}}{dt} + v_{it}  + \frac{dn_{i,t-1}}{dt} \, v_{it} - E_{it} - \sum_{j, s} J_{ijs} \, dn_{j,t-s} \right) \\
&- \tilde{n}_{it} dn_{it} + \ln \big(1 + f(v_{it})\, dt\left(\exp(\tilde{n}_{it}) -1\right) \big) \Bigg).
\end{aligned} \end{equation}
\end{widetext}
The measures are $D\tilde{\bm{n}} = \prod_{i,t} \frac{d\tilde{n}_{it}}{2 \pi i}$ and $D \tilde{\bm{v}} = \prod_{i, t} \frac{d\tilde{v}_{it}}{2 \pi i}$. The integrals over the response variables, $\tilde{\bm{n}}$ and $\tilde{\bm{v}}$, are along the imaginary axis.
The logarithmic term in the exponent is the cumulant-generating function of the Bernoulli spikes. 
Galves \& L\H{o}cherbach proved the existence and uniqueness of stationary densities for the discrete-time model with strictly positive intensity function~\cite{galves_infinite_2013}.

We next take a continuous time limit, $dt \rightarrow 0, T \rightarrow \infty$, with their product fixed. This defines the functional integration measures $\mathcal{D}$.
With $dt \ll 1$, we expand the natural logarithm in its Taylor series around 1: $\ln \big(1 + f(v_{it})\, dt\left(\exp(\tilde{n}_{it}) -1\right) \big) = f(v_{it})\, dt\left(\exp(\tilde{n}_{it}) -1\right) + \mathcal{O}\big((dt)^2\big)$. 
This yields Eq.~\ref{eq:lif_action}, with the infinitesimal shift in the reset term in Eq.~\ref{eq:lif}.

\section{Absolute refractory period} \label{app:absolute_refractory}
With an absolute refractory period of $T_r$ time steps in the discrete-time dynamics, during which the membrane voltage is clamped within $\mathcal{O}(dt)$ of 0, the joint density of the spike trains and membrane voltages instead obeys
\begin{widetext}
\begin{equation} \begin{aligned}
p(\bm{v}, \bm{n}) = \int D\tilde{\bm{v}} \int D \tilde{\bm{n}} \; \exp \Bigg(& \sum_{i, t} -\tilde{v}_{it} \left(\frac{dv_{it}}{dt} + v_{it}  + \sum_{s=1}^{T_r} \frac{dn_{i, t-s}}{dt} \, v_{it} - E_{it} - \sum_{j, s} J_{ijs} \, dn_{j,t-s} \right) \\
&- \tilde{n}_{it} dn_{it} + \ln \big(1 + f(v_{it})\, dt\left(\exp(\tilde{n}_{it}) -1\right) \big) \Bigg).
\end{aligned} \end{equation}
\end{widetext}
This presents some complication in the continuous-time limit: the refractory term diverges when written as a convolution. 

One alternative would be to incorporate a strong, negative self-coupling in diagonal elements of $\bm{J}$. While not strictly an absolute refractory period, this may mimic its effects. This would affect the definition of the mean field theory and propagators, but would not give rise to new types of fluctuation correction (no new vertices; Appendix~\ref{app:feynman_rules}).

Another alternative is consider an absolute refractory period in which the membrane voltage is not clamped at the reset voltage. Rather, we can require that 
\begin{enumerate}
\item during the absolute refractory period, the spike probability is 0 and
\item at the end of the absolute refractory period, the membrane voltage is reset to the reset voltage.
\end{enumerate}
This yields the discrete-time density
\begin{widetext}
\begin{equation} \begin{aligned}
p(\bm{v}, \bm{n}) =& \int D\tilde{\bm{v}} \int D \tilde{\bm{n}} \; \exp \Bigg(\sum_{i, t} -\tilde{v}_{it} \left(\frac{dv_{it}}{dt} + v_{it}  + \frac{dn_{i, t-{T_r}}}{dt} \, v_{it} - E_{it} - \sum_{j, s} J_{ijs} \, dn_{j,t-s} \right) \\
&\mkern 155mu - \tilde{n}_{it} dn_{it} + \ln \left(1 + f(v_{it})\, dt \left(\exp(\tilde{n}_{it}) -1\right) \left(1 - \sum_{s=1}^{T_r} dn_{i, t-s}\right) \right) \Bigg), 
\end{aligned} \end{equation}
\end{widetext}
from which a continuum limit can be taken straightforwardly, yielding a spike reset term $\dot{n}_i(t-\tau_r) \, v_i(t)$ and spike intensity $f(v_i) \left(1 - \dot{n}_i \ast B \right)$, with the rectangular function $B(t) = \theta(t) - \theta(t - \tau_r)$ and refractory period $\tau_r = T_r \, dt$.
This introduces a new state-dependence to the intensity, which would give rise to new types of fluctuation correction.

\section{Connectivity-averaged density} \label{app:mft_homog}
To examine the interaction between synaptic connectivity, subthreshold dynamics, and stochastic spike emission in shaping network activity, we will average the partition functional for the activity (equivalently, average the moment generating functional) over the synaptic connectivity. This is a standard exercise in statistical field theory~\cite{helias_statistical_2020}, relying on the assumption that the system is self-averaging with respect to the connectivity: that is, that the average over realizations of $\bm{J}$ will give us an accurate description of a single large system.
The connectivity-averaged partition functional is
\begin{equation} \begin{aligned}
Z =&  \int \mathcal{D}\bm{v} \int \mathcal{D}\dot{\bm{n}}  \int \mathcal{D}\tilde{\bm{v}} \int \mathcal{D}\tilde{\bm{n}}  \int D\bm{J} \; p[v, \dot{n}, \tilde{v}, \tilde{n} \vert \bm{J}] \; p(\bm{J}).
\end{aligned} \end{equation}
Since the action $S$ is linear in $\bm{J}$, a cumulant generating function for $\bm{J}$ appears in $Z$:
\begin{widetext}
\begin{equation} \begin{aligned}
Z =& \int \mathcal{D}\bm{v} \int \mathcal{D}\dot{\bm{n}} \int \mathcal{D}\tilde{\bm{v}} \int \mathcal{D}\tilde{\bm{n}} \; \exp -\Bigg( \tilde{\bm v}^T \left(\partial_t \bm{v} +{ \bm v} - {\bm E} + \dot{\bm n} {\bm v} \right) + \tilde{\bm n}^T \dot{\bm n} - \left( \exp (\tilde{\bm n})-1 \right)^T {\bm f} \Bigg) \int D\bm{J} \; p(\bm{J}) \exp \left(\tilde{\bm v}^T {\bm J} \ast \dot{\bm n}  \right)  \\
=&  \int \mathcal{D}\bm{v} \int \mathcal{D}\dot{\bm{n}}  \int \mathcal{D}\tilde{\bm{v}} \int \mathcal{D}\tilde{\bm{n}} \; \exp -\Bigg( \tilde{\bm v}^T \left(\partial_t \bm{v} +{ \bm v} - {\bm E} + \dot{\bm n} {\bm v}  \right) - W_{\bm{J}}(\tilde{\bm v}^T \dot{\bm n}) + \tilde{\bm n}^T \dot{\bm n} - \left( \exp (\tilde{\bm n})-1 \right)^T {\bm f} \Bigg) ,
\end{aligned} \end{equation}
\end{widetext}
where
\begin{equation} \begin{aligned}
W_{\bm{J}}(\tilde{\bm v}^T \dot{\bm n}) =& \ln \int D\bm{J} \; p(\bm{J}) \exp \Bigg( \int dt \int ds \sum_{i, j} \\
&\tilde{v}_i(t) \, J_{ij}(s) \, \dot{n}_j(t-s) \Bigg) .
\end{aligned} \end{equation}
Due to this, each cumulant of $\bm{J}$ gives rise to a corresponding cumulant in the connectivity-averaged partition functional for the activity. 
We have overloaded notation here, writing $p(\bm{J})$ for the distribution of the synaptic weight matrix while also letting $\bm{J}$ be a function of the time lag. 
This notation assumes that $J_{ij}(s) = J_{ij} \, G_{ij}(s)$ for some matrix of unit-norm kernels ${\bm G}$, which we leave implicit.
The connectivity gives rise to an effective noise in the membrane voltage. 
Each cumulant of the connectivity gives rise to a cumulant of the same order in the effective noise. 

For example, consider an Erd\H{o}s-R\'enyi network with connection probability $p$ and synaptic weight $J$ for the non-zero connections, with $J(s) = J \, \delta(s)$. 
The distribution $p(\bm{J})$ factorizes over the weights; the cumulant generating functional for an individual synaptic weight is
\begin{equation}
W_J(x) = \ln \left(1 + p\left(-1 + \exp \left(J x\right) \right) \right)
\end{equation}
and cumulants of the synaptic weights $J_{ij}$ obey the recursion relation
\begin{equation} \begin{aligned}
\langle J_{ij} \rangle =& J p \\
\llrrangle{J_{ij}^n} =& J p(1-p) \frac{d}{dp} \llrrangle{J_{ij}^{n-1}}, \; n \geq 2.
\end{aligned} \end{equation}
If the connection probability $p$ and weight $J$ are both of order 1, the synaptic weights will have non-negligible cumulants of all orders. 
If the synaptic weights are of order one and the connectivity sparse, $p \sim 1/N$, the cumulants are $\llrrangle{J_{ij}^n} = J^n p + \mathcal{O}(1/N^2)$. 
If $J > 1$, higher cumulants of the connectivity will dominate, giving rise to higher-order cumulants in the effective noise of the membrane voltage. 
In contrast, if $J \sim 1/N$ and $p \sim 1$ then $\llrrangle{J_{ij}^n} \sim N^{-n}$ so in a large network, the first cumulant of the connectivity dominates.

Here, we consider that simple case where $\bm{J}$ has only a first cumulant. 
Let $\langle J_{ij}(s) \rangle_{\bm{J}} = J(s) / N$.  The average over the connectivity yields
\begin{widetext}
\begin{equation} \begin{aligned}
Z =& \int \mathcal{D} \bm{v} \int \mathcal{D} \dot{\bm{n}} \int \mathcal{D} \tilde{\bm{v}}  \int \mathcal{D} \tilde{\bm{n}} \; \exp \sum_i 
\Bigg(-\tilde{v}_i^T \left(\partial_t v_i + v_i + \dot{n}_i v_i - E_i - \frac{1}{N} J \ast \sum_j \dot{n}_j \right) \\
&- \tilde{n}_i^T \dot{n}_i + \left(e^{\tilde{n}_i}-1 \right)^T f(v_i)  \Bigg) .
\end{aligned} \end{equation}
\end{widetext}
(Here, $\ast$ represents scalar temporal convolution.)
We would like to examine this partition functional in the limit of a large network.
Let $R = \frac{1}{N} \sum_j J \ast \dot{n}_j$; we will enforce this by integrating against $\delta\left(NR - J \ast \sum_j \dot{n}_j \right)$.
With the Fourier representation of that delta function, we have a generating functional for the auxiliary fields $R, \tilde{R}$:
\begin{widetext}
\begin{equation} \begin{aligned}
Z[k, \tilde{k}] =& \int \mathcal{D}R \int \mathcal{D} \tilde{R} \; \exp \left(-N \tilde{R}^T R + \sum_i \ln Z_i[R, \tilde{R}] + \tilde{k}^T R + k^T \tilde{R} \right), \\
Z_i[R, \tilde{R}] =& \int \mathcal{D} v_i  \int \mathcal{D} \dot{n}_i \int \mathcal{D} \tilde{v}_i   \int \mathcal{D} \tilde{n}_i \; \exp \Bigg(-\tilde{v}_i^T \left( \partial_t v_i +v_i - E -R  + \dot{n}_i v_i \right) + \tilde{R}^T \left(J \ast \dot{n}_i \right) \\
&- \tilde{n}_i^T \dot{n}_i + \left(e^{\tilde{n}_i}-1 \right)^T f(v_i) \Bigg).
\end{aligned} \end{equation}
\end{widetext}
Note that the generating function for the neural dynamics factorizes over the neurons; $Z_i[R, \tilde{R}]$ does not contain any other indices. 
So, we will drop the neuron indices and write $N \ln Z[R, \tilde{R}]$ instead of $\sum_i \ln Z_i[R, \tilde{R}]$. 
For large $N$, we evaluate the integrals over the auxiliary fields $R, \tilde{R}$ by a saddle point approximation. The saddle point equations are
\begin{equation} \begin{aligned}
0 &= -N R^* + N\frac{\partial \ln Z[R, \tilde{R}]}{\partial \tilde{R}}\vert_R^* \leftrightarrow R^* = J \ast \langle \dot{n} \rangle, \\
0 =& -N \tilde{R}^* + N\frac{\partial \ln Z[R, \tilde{R}]}{\partial R}\vert_R^* \leftrightarrow \tilde{R}^* = \langle \tilde{v} \rangle = 0.
\end{aligned} \end{equation}
Here, $\langle \dot{n} \rangle(t)$ is the population-averaged firing rate.
Inserting these saddle-point solutions yields the partition functional, Eq.~\ref{eq:saddle_point_Z_1pop_homog}.

\section{Perturbative expansion} \label{app:perturb}
If fluctuations or nonlinearities are weak, a perturbative expansion around the mean field theory can provide accurate estimates of fluctuation effects.
For ease of notation, let $\bm{x} = (v, \dot{n})^T$ and $\tilde{\bm{x}} = (\tilde{v}, \tilde{n})^T$. We expand $\bm{x}$ around a background field,
\begin{equation} \label{eq:shift}
\bm{x} = \bar{\bm{x}} + \delta \bm{x}, \; \tilde{\bm{x}} = \tilde{\bm{x}}^*  + \delta \tilde{\bm{x}},
\end{equation}
and collect terms up to linear order in the fluctuations in the free action $S_0$, with higher order terms in the interacting part of the action $S_V$:
\begin{equation} \begin{aligned} \label{eq:action_separated}
S =& S_0 + S_V \\
S_0 =& \tilde{v}^T \left( \partial_t \bar{v} + \bar{v} + \bar{v}\bar{n} - E - J \ast \langle \dot{n} \rangle \right) + \tilde{n}^T \left(\bar{n} - f(\bar{v}) \right)\\
&+ \tilde{v}^T \left(\partial_t + 1 + \bar{n} \right) \delta v + \tilde{v}^T \bar{v} \, \delta n + \tilde n^T \delta n - \tilde{n}^T f^{(1)} \, \delta v, \\
S_V =& \tilde{v}^T \, \delta n \, \delta v - \sum_{p=2}^\infty \frac{\tilde{n}^p}{p!} f(\bar{v}) - \sum_{\substack{p, q=1 \\ p+q>2}}^\infty \frac{\tilde{n}^p}{p!} \frac{f^{(q)}}{q!} (\delta v)^q .
\end{aligned} \end{equation}
(We should also expand the response variable $\tilde{\bm x}$ around a background field; we skip that here since in Appendix~\ref{app:effective_action} we will constrain the background fields to be the mean trajectories, and the mean of response variables is 0.)
A joint moment of $v, \dot{n}$ is
\begin{equation} \begin{aligned}
\left \langle \prod_{i=1}^a \dot{n}(t_i) \prod_{j=1}^b v(t_j) \right \rangle =& \int \mathcal{D} \delta \bm{x} \int \mathcal{D} \tilde{\bm{x}} \; \prod_{i=1}^a \prod_{j=1}^b \\
& \dot{n}_i  v_j \exp \Big(-S_0 - S_V \Big) .
\end{aligned} \end{equation}
Expanding $\exp \left(-S_V\right)$ in a functional Taylor series around a solution to the mean field theory yields an expansion of the moment in terms of Gaussian integrals with respect to the free density $\exp \left(-S_0\right)$. Due to Wick's theorem, these integrals yield products of the propagators, $\bar{\bm \Delta}$. These expansions can be efficiently organized diagrammatically.  

\subsection{Feynman rules} \label{app:feynman_rules}
Here we give the Feynman rules for a perturbative expansion of statistics of the population-averaged system, Eq.~\ref{eq:saddle_point_Z_ei_homog}, around the mean field theory.
This provides a graphical algorithm for computing arbitrary cumulants of the spike train $\dot{n}$ or membrane voltage $v$. 
Moments can be composed from the cumulants by the appropriate Bell polynomials.
We give the rules in the temporal frequency domain, for an expansion around a stationary point.
Each cumulant can be decomposed into a sum of terms, each represented by a connected diagram. 
Those diagrams are composed of the vertices and edges in Tables~\ref{table:vertices} and~\ref{table:edges} (Feynman diagrams generated with~\cite{ellis_tikz-feynman_2017}).
{\renewcommand{\arraystretch}{2} 
\begin{table}[ht!]
\begin{tabular}{| c | c | c | c |} \hline 
Vertex & Factor & \makecell{ In-degree \\ ($\delta v, \; \delta n$)} & \makecell{ Out-degree \\ ($\tilde{v}, \; \tilde{n}$)} \\ \hline
$\feynmandiagram{a[dot]};$ & $f(\bar{v})$ & (0, 0) & $(0, \; \geq 2)$  \\ \hline
$\feynmandiagram[]{a[empty dot]};$ & $\frac{f^{(q)}}{ q! } \, \delta \Big(\sum_{i=1}^{K_{\mathrm{in}}} \omega_{i} - \sum_{j=1}^{K_{\mathrm{out}}} \omega_{j} \Big)$ & $ (\geq 1, \; 0)$ & $(0, \; q), \; q\geq 1$ \\ \hline 
$\feynmandiagram[]{a[empty dot]};$ & $ - \delta \Big(\sum_{i=1}^{K_{\mathrm{in}}} \omega_{i} - \sum_{j=1}^{K_{\mathrm{out}}} \omega_{j} \Big) $ & (1, 1) & (1, 0) \\ \hline  
\end{tabular} 
\caption{\label{table:vertices}Vertices corresponding to the interacting action, $S_V$, in Eq.~\ref{eq:action_separated}. $f^{(q)}$ is the $q$th derivative of the intensity function $f$, evaluated at the expansion point $\bar{v}$. The intensity function vertex, $f^{(q)} / q!$, also has the constraint that the sum of its in and out-degrees must be at least three since the linear and bilinear terms in $(\tilde{n}, \delta v)$ went in to the definitions of the background field and the propagators (Eq.~\ref{eq:action_separated}).
}
\end{table}
}

{ \renewcommand{\arraystretch}{2}
\begin{table}
\begin{tabular}{| c | c | c |} \hline 
Edge & Propagator & Factor \\ \hline
\feynmandiagram[horizontal=a to b, inline=(a.base)]{a --[] b}; &  $\bar{\Delta}_{ n, \tilde{n}}(\omega) $ & $\left(1 + \bar{n} + i \omega \right) / \left(1 + \bar{n} + f^{(1)}\bar{v} + i \omega\right)$ \\[2\baselineskip] \hline
\feynmandiagram[horizontal=a to b, inline=(a.base)]{a --[photon] b}; & $\bar{\Delta}_{ v, \tilde{n}}(\omega)$ & $ -\bar{v} / \left(1 + \bar{n} + f^{(1)}\bar{v} + i \omega\right)$  \\ \hline
\feynmandiagram[horizontal=a to b, inline=(a.base)]{a --[gluon] b}; & $ \bar{\Delta}_{ n, \tilde{v}}(\omega) $ &  $f^{(1)} / \left(1 + \bar{n} + f^{(1)}\bar{v} + i \omega\right) $  \\ \hline
\feynmandiagram[horizontal=a to b, inline=(a.base)]{a --[scalar] b}; & $ \bar{\Delta}_{ v, \tilde{v}}(\omega)$ & $ 1 / \left(1 + \bar{n} + f^{(1)}\bar{v} + i \omega\right)$ \\ \hline
\end{tabular}
\caption{\label{table:edges} Edges corresponding to the components of the propagator from $S_0$ in Eq.~\ref{eq:action_separated}. Each measures the linear response of one configuration variable to a perturbation of another. For example, $\bar{\Delta}_{ v, \tilde{n}}$ measures the linear response of the voltage to a spike fluctuation. 
}
\end{table}
}
The source vertex $\feynmandiagram{a[dot]};$ emits factors of the response variable $\tilde{n}$ corresponding to spike fluctuations. 
Each internal vertex, $\feynmandiagram{a[empty dot]};$, receives configuration variables $\delta n, \; \delta v$ and emits response variables, $\tilde{n}, \; \tilde{v}$. 
In any connected diagram, each pair of vertices will be linked by at least one pair of configuration and response variables, e.g., ($\delta v, \tilde{n}$). 
Due to Wick's theorem, each pair of configuration and response variables is replaced by the corresponding propagator edge. 
For example, the pair $\delta v, \tilde{n}$ gives rise to the propagator $\bar{\Delta}_{ v, \tilde{n}}$.
To calculate the joint cumulant $\llrrangle{\dot{n}^a v^b}(\omega_1, \ldots, \omega_{a+b-1})$:
\begin{enumerate}
\item Place an external vertex for each of the $a+b$ factors of $\dot{n}$ and $v$.
\item Using the internal vertices and edges in Tables~\ref{table:vertices},~\ref{table:edges}, construct all connected graphs such that each external vertex has one incoming propagator edge. Each edge has its own frequency variable, $\omega_i$. 
\item To evaluate a diagram, multiply the factors of every edge and vertex together. Additionally, the sum of external frequencies (those on the external vertices' incoming edges) is zero: also multiply by $\delta \left(\sum_{i=1}^{a+b} \omega_{i} \right)$. Finally, integrate over all of the internal frequencies: for each internal frequency $\omega_i$, integrate $\int d\omega_i / 2\pi$. 
\item Evaluate each connected diagram constructed in (2), and add the contributions of the diagrams.
\end{enumerate}

We perform the integrals over internal frequencies analytically using the residue theorem. For a thorough introduction see e.g.,~\cite{zinn-justin_quantum_2002, helias_statistical_2020}.
For an introduction to diagrammatic methods in the Poisson generalized linear model without resets (no self-coupling) see~\cite{ocker_linking_2017}. 
See~\cite{kordovan_spike_2020} for detailed analytical calculations of the integrals over internal frequencies in that model.

Eqs.~\ref{eq:lif_feynman_1loop},~\ref{eq:feynman_1loop_1pop} are self-consistent one-loop equations of motion for the mean voltage and rate.
The approximate joint cumulants appearing in them can also be calculated using the perturbative Feynman rules above; each is given by a tree diagram with the same two edges as in the loop diagram.

The corresponding perturbative corrections to the mean field values of the voltage and rate are given by one-loop tadpole diagrams. For example, the perturbative one-loop correction to the mean field voltage corresponding to Eq.~\ref{eq:bubble_reset} is $\vcenter{ \hbox{
 \feynmandiagram[scale=0.5][horizontal=a to b, inline=(b.base)]{
a --[dashed] b[empty dot] --[half left, photon] c[dot] --[half left] b,
 };  }} 
 $. 
 The internal vertex $\feynmandiagram{a[empty dot]};$ in this diagram carries a factor of $-1$. This corresponds to the sign this diagram appears with in Eqs.~\ref{eq:lif_feynman_1loop},~\ref{eq:feynman_1loop_1pop}, opposite the sign in front of the second loop diagram. There is also a perturbative one-loop correction to the mean field voltage arising from the intensity vertex, 
 $\vcenter{ \hbox{
 \feynmandiagram[scale=0.5][horizontal=a to b, inline=(b.base)]{
a --[photon] b[empty dot] --[half left, photon] c[dot] --[half left, photon] b,
 };  }} 
 $. Both nonlinearities also give rise to perturbative corrections to the mean field rate. This is a difference with the self-consistent approach of the main text, where only one one-loop correction arises in each equation of motion (Appendix~\ref{app:effective_action}).

\section{Effective action} \label{app:effective_action}
Here we briefly derive the effective action. For a more detailed presentation see e.g.,~\cite{zinn-justin_quantum_2002} Ch. 7 or~\cite{helias_statistical_2020} Ch. 11-14.
For ease of notation, let $\bm{x} = (v, \dot{n})^T$ and $\tilde{\bm{x}} = (\tilde{v}, \tilde{n})^T$. 
The cumulant-generating functional is
\begin{equation} \label{eq:cgf}
\exp W[\bm{j}, \tilde{\bm{j}}] = \int \mathcal{D}\bm{x} \int \mathcal{D}\tilde{\bm{x}} \; \exp \frac{1}{h} \left( -S[\bm{x}, \tilde{\bm{x}}] + \tilde{\bm j}^T \bm{x} + \bm{j}^T \tilde{\bm{x}} \right).
\end{equation}
We have introduced a scale $h$ into the exponent on the right-hand side. For physical calculations we will set $h = 1$. (Here, $h$ has no relation to that used in Section~\ref{sec:isn}.)
We expand $\bm{x}$ around a background field $\bar{\bm{x}}$ (Eq.~\ref{eq:shift}) and similarly for the response variable, $\tilde{\bm{x}} = \tilde{\bm x}^* + \delta \tilde{\bm x}$.
This yields
\begin{equation} \begin{aligned} \label{eq:shifted_cumulant_generator}
&\exp \frac{1}{h}\left( hW[\bm{j}, \tilde{\bm{j}}] - \tilde{\bm j}^T \bar{\bm{x}} - \bm{j}^T \tilde{\bm{x}}^* \right)  \\
=& \int \mathcal{D} \delta \bm{x} \int \mathcal{D} \delta \tilde{\bm{x}} \; \exp  \frac{1}{h} \Big( -S + \tilde{\bm j}^T \delta \bm{x} + \bm{j}^T \delta \tilde{\bm{x}} \Big).
\end{aligned} \end{equation}
We now require that our background field be the mean: $\bar{\bm{x}} = \langle \bm{x} \rangle$ so that 
\begin{equation} \begin{aligned}
0 =& \langle \delta x_i \rangle = \frac{\partial}{\partial \tilde{j}_i} \exp \frac{1}{h} \left( h W[\bm{j}, \tilde{\bm{j}}]  - \tilde{\bm j}^T \bar{\bm{x}}  - \bm{j}^T \tilde{\bm{x}}^* \right)
\end{aligned} \end{equation}
and similarly, we require $\tilde{\bm{x}}^* = \langle \tilde{\bm{x}} \rangle$.
These requirements can only be satisfied at a stationary point of the exponent:
\begin{equation} \begin{aligned}
0 =& \frac{\partial}{\partial j_i} \left( h W[\bm{j}, \tilde{\bm{j}}] - \tilde{\bm j}^T \bar{\bm x}  - \bm{j}^T \tilde{\bm{x}}^* \right) \\
0 =& \frac{\partial}{\partial \tilde{j}_i} \left( h W[\bm{j}, \tilde{\bm{j}}] - \tilde{\bm j}^T \bar{\bm x}  - \bm{j}^T \tilde{\bm{x}}^* \right),
\end{aligned} \end{equation}
which defines a Legendre transform from $-h W$ to the effective action $\Gamma$:
\begin{equation}
\Gamma[\bar{\bm{x}}, \tilde{\bm{x}}^*] = \sup_{\bm{j}, \tilde{\bm{j}}}  \tilde{\bm j}^T \bar{\bm x}  + \bm{j}^T \tilde{\bm{x}}^* - h W[\bm{j}, \tilde{\bm{j}}].
\end{equation}
Substituting the effective action in Eq.~\ref{eq:shifted_cumulant_generator} yields
\begin{equation} \begin{aligned} \label{eq:effective_action_fluctuations}
&\exp \frac{1}{h} \left( -\Gamma[\bar{\bm{x}}, \tilde{\bm{x}}^*] + S[\bar{\bm{x}}, \tilde{\bm{x}}^*] \right) \\
=& \int \mathcal{D}\delta \bm{x} \int \mathcal{D}\delta \tilde{\bm{x}} \; \exp \frac{1}{h} \left( -S[\delta \bm{x}, \delta \tilde{\bm{x}}] + \tilde{\bm j}^T \delta \bm{x} + \bm{j}^T \delta \tilde{\bm{x}} \right) ,
\end{aligned} \end{equation}
where $S[\bar{\bm{x}}, \tilde{\bm{x}}^*]$ contains the terms in the $S$ that depend only on $\bar{\bm{x}}, \tilde{\bm{x}}^*$ and not $\delta {\bm x}, \delta \tilde{\bm x}$ and $S[\delta \bm{x}, \delta \tilde{\bm{x}}]$ contains the remaining terms of $S$.
This has the form of a generating functional for $\bar{\bm{x}}, \tilde{\bm{x}}^*$. 
The mean is a stationary point of $\Gamma$; it obeys the equations of motion
\begin{equation} \begin{aligned}
0 =& \frac{\partial }{\partial \, \bar{x}_i} \Gamma \\
0 =& \frac{\partial }{\partial \, \tilde{x}_i^*} \Gamma .
\end{aligned} \end{equation}
The loop expansion for the effective action is a diagrammatic equivalent of the saddle point expansion of the integrals over $\delta \bm{x}, \delta \tilde{\bm{x}}$ in Eq.~\ref{eq:effective_action_fluctuations}, without requiring that $h$ be a bona fide small parameter~\cite{coleman_radiative_1973, buice_systematic_2010}. 
The diagrams contributing to the equations of motion for $\bar{\bm{x}}, \tilde{\bm{x}}^*$ are one-line-irreducible vacuum diagrams (those that cannot be disconnected by cutting one edge; see e.g.,~\cite{helias_statistical_2020}, Ch. 11.4, 13.3). Only those diagrams with a vertex carrying the appropriate factor of $\tilde{x}_i^*$ will contribute to the equation of motion $0 = \frac{\partial }{\partial \, \tilde{x}_i^*} \Gamma $ (and similarly for $\partial / \partial \bar{x}_i$) which is why the one-loop equations of motion each have only one loop correction.
\\

\section{Connectivity average for the linear-reset model} \label{app:linear_reset}
The connectivity-averaged action for the linear-reset model is
\begin{equation} \begin{aligned} 
S[v, \dot{n}, \tilde{v}, \tilde{n}] =& \tilde{v}^T \left( \partial_t v + v + r\dot{n} - E - J \langle \dot{n} \rangle \right) \\
&+ \tilde{n}^T \dot{n} - \left(e^{\tilde{n}} -1\right)^T f(v)
\end{aligned} \end{equation}
and expanding around a solution to the mean-field theory $\partial_t \bar{v} =-\bar{v} + (J - r) \langle \dot{n} \rangle + E$ yields the free and interacting actions
\begin{equation} \begin{aligned} \label{eq:action_linear_reset_1pop_mft_expansion}
S_0 =& \tilde{v}^T \left( \partial_t  \bar{v} + \bar{v} + r \bar{n} - E - J \ast \langle \dot{n} \rangle \right) + \tilde{n}^T \left( \bar{n} - f(\bar{v}) \right) \\
&+\tilde{v}^T \left(\partial_t + 1 \right) \delta v + \tilde{v}^T r \, \delta n + \tilde n^T \delta n - \tilde{n}^T f^{(1)} \, \delta v, \\
S_V =& - \sum_{p=2}^\infty \frac{\tilde{n}^p}{p!} \bar{f} - \sum_{\substack{p, q=1 \\ p+q>2}}^\infty \frac{\tilde{n}^p}{p!} \frac{f^{(q)}}{q!} (\delta v)^q.
\end{aligned} \end{equation}

The components of the propagator for this model are given in Table~\ref{table:edges_linear_reset}. It has the same source and intensity vertices as the stochastic LIF model (the first two entries in Table~\ref{table:vertices}), but lacks the reset vertex.
{ \renewcommand{\arraystretch}{2}
\begin{table}[ht!]
\begin{tabular}{| c | c | c |} \hline 
Edge & Propagator & Factor \\ \hline
\feynmandiagram[horizontal=a to b, inline=(a.base)]{a --[] b}; &  $\bar{\Delta}_{ n, \tilde{n}}(\omega) $ & $\left(1 + i \omega \right) / \left(1 + f^{(1)} r + i \omega\right)$ \\[2\baselineskip] \hline
\feynmandiagram[horizontal=a to b, inline=(a.base)]{a --[photon] b}; & $\bar{\Delta}_{ v, \tilde{n}}(\omega)$ & $- r / \left(1 + f^{(1)} r + i \omega\right)$  \\ \hline
\feynmandiagram[horizontal=a to b, inline=(a.base)]{a --[gluon] b}; & $ \bar{\Delta}_{ n, \tilde{v}}(\omega) $ & $ f^{(1)} / \left(1 + f^{(1)} r+ i \omega\right) $  \\ \hline
\feynmandiagram[horizontal=a to b, inline=(a.base)]{a --[scalar] b}; & $ \bar{\Delta}_{ v, \tilde{v}}(\omega)$ & $ 1 / \left(1 + f^{(1)} r + i \omega\right)$ \\ \hline
\end{tabular}
\caption{\label{table:edges_linear_reset} Components of the propagator from $S_0$ in Eq.~\ref{eq:action_linear_reset_1pop_mft_expansion}.}
\end{table}
}

\end{document}